\documentclass{aa}  

\usepackage{graphicx}
\usepackage{txfonts}
\usepackage[colorlinks = true,
            linkcolor = blue,
            urlcolor  = black,
            citecolor = cyan,
            anchorcolor = black]{hyperref}
\newcommand{\ppxf}{pPXF}
\newcommand {\ha} {H$\alpha$}
\newcommand {\hb} {H$\beta$}
\newcommand{\hei}{[He\,\textsc{I}]}
\newcommand{\oiii}{[O\,\textsc{iii}]}
\newcommand{\nii}{[N\,\textsc{ii}]}
\newcommand{\sii}{[S\,\textsc{ii}]}
\newcommand{\oi}{[O\,\textsc{i}]}

\newcommand{\fexxvi}{Fe\,\textsc{xxvi}}

\newcommand{\feii}{[Fe\,\textsc{ii}]}
\newcommand{\hi}{H\,\textsc{i}}
\newcommand{\nai}{Na\,\textsc{i}}

\newcommand {\kms} {\,{\rm km\,s}^{-1}}
\newcommand {\ergs} {\,{\rm erg\,s}^{-1}}
\newcommand {\erg} {\,{\rm erg}}
\newcommand {\pc} {\,{\rm pc}}

\newcommand {\kpc} {\,{\rm kpc}}
\newcommand {\Mpc} {\,{\rm Mpc}}
\newcommand {\cmmq}{\,{\rm cm^{-2}}}
\newcommand {\cmmc}{\,{\rm cm^{-3}}}

\newcommand {\kmsMpc} {\,{\rm km\,s}^{-1}\,{\rm \Mpc}^{-1}}

\newcommand{\lala}{\lambda\lambda}

\newcommand {\lsun}{\,{\rm L}_\odot}
\newcommand {\msun}{\,{\rm M}_\odot}

\newcommand{\Myr}{\,{\rm Myr}}

\newcommand{\K}{\,{\rm K}}

\newcommand {\msunyr}{\,{{\rm M}_\odot\,\rm yr}^{-1}}

\newcommand{\avg}[1]{\left< #1 \right>} % for average

\def\apjl{ApJ}%
\def\apjs{ApJS}%
\def\ao{Appl.~Opt.}%
\def\aap{A\&A}%
\begin{document} 

\renewcommand{\figureautorefname}{Figure\!}
\title{Galaxy-scale ionised winds driven by ultra-fast outflows in two nearby quasars}
\titlerunning{Ionised winds in QSOs hosting UFOs}
\authorrunning{A. Marasco et al.}
%   \subtitle{}

   \author{A. Marasco\inst{1},
          G. Cresci\inst{1},
          E. Nardini\inst{1,2},
          F. Mannucci\inst{1},
          A. Marconi\inst{1,2},
          P. Tozzi\inst{1},
          G. Tozzi\inst{1,2},
          A. Amiri\inst{1,2},
          G. Venturi\inst{3},
          E. Piconcelli\inst{4},
          G. Lanzuisi\inst{5},
          F. Tombesi\inst{6,4,7,8},
          M. Mingozzi\inst{9},
          M. Perna\inst{10,1},
          S. Carniani\inst{11},
          M. Brusa\inst{12,5},
          S. di Serego Alighieri\inst{1}
          }

\institute{INAF - Osservatorio Astrofisico di Arcetri, Largo E. Fermi 5, 50127, Firenze, Italy\\
\email{antonino.marasco@inaf.it}
\and
Dipartimento di Fisica e Astronomia, Università di Firenze, Via G. Sansone 1, 50019, Sesto Fiorentino (Firenze), Italy
\and
Instituto de Astrof{\'{i}}sica, Pontificia Universidad Cat{\'{o}}lica de Chile, Avda. Vicu{\~{n}}a Mackenna 4860, 8970117, Macul, Santiago, Chile \and
Osservatorio Astronomico di Roma - INAF, via Frascati 33, 00040 Monte Porzio Catone, Italy
\and
INAF - Osservatorio di Astrofisica e Scienza dello Spazio di Bologna, via Gobetti 93/3, 40129 Bologna, Italy
\and
Dipartimento di Fisica, Univerisità di Roma Tor Vergata, via della Ricerca Scientifica 1, I-00133 Roma, Italy
\and
Department of Astronomy, University of Maryland, College Park, MD 20742, USA
\and
X-ray Astrophysics Laboratory, NASA/Goddard Space Flight Center, Greenbelt, MD 20771, USA
\and
INAF - Osservatorio Astronomico di Padova, Vicolo dell’Osservatorio 5, I-35122 Padova, Italy
\and
Centro de Astrobiología (CSIC-INTA), Departamento de Astrofísica, Cra. de Ajalvir Km. 4, 28850, Torrejón de Ardoz, Madrid, Spain
\and
Scuola Normale Superiore, Piazza dei Cavalieri 7, I-56126 Pisa, Italy
\and
Dipartimento di Fisica e Astronomia, Alma Mater Studiorum Università di Bologna, via Gobetti 93/2, 40129 Bologna, Italy 
}

   \date{Received ; accepted}

% \abstract{}{}{}{}{} 
% 5 {} token are mandatory
 
\abstract
{
We use MUSE adaptive optics (AO) data in Narrow Field Mode to study the properties of the ionised gas in MR\,2251-178 and PG\,1126-041, two nearby ($z\simeq0.06$) bright quasars hosting sub-pc scale Ultra Fast Outflows (UFOs) detected in the X-ray band.
We decompose the optical emission from diffuse gas into a low- and a high-velocity components.
The former is characterised by a clean, regular velocity field and a low ($\sim80\kms$) velocity dispersion.
It traces regularly rotating gas in PG\,1126-041, while in MR\,2251-178 it is possibly associated to tidal debris from a recent merger or flyby.
The other component is found to be extended up to a few kpc from the nuclei, and shows a high ($\sim800\kms$) velocity dispersion and a blue-shifted mean velocity, as expected from AGN-driven outflows.
We estimate mass outflow rates up to a few $\msunyr$ and kinetic efficiencies $L_{\rm KIN}/L_{\rm BOL}$ between $1-4\times10^{-4}$, in line with those of galaxies hosting AGNs of similar luminosity.
The momentum rates of these ionised outflows are comparable to those measured for the UFOs at sub-pc scales, consistent with a momentum-driven wind propagation. Pure energy-driven winds are excluded unless about $100\times$ additional momentum is locked in massive molecular winds.
By comparing the outflow properties of our sources with those of a small sample of well-studied QSOs hosting UFOs from the literature, we find that winds seem to systematically lie either in a momentum-driven or in an energy-driven regime, indicating that these two theoretical models bracket very well the physics of AGN-driven winds.}% and that possible transitions from one regime to the other must occur on short timescales.}

\keywords{galaxies: ISM -- quasars: individual: MR\,2251-178 -- quasars: individual: PG\,1126-041 -- ISM: jets and outflows -- techniques: imaging spectroscopy}
\maketitle
%
%________________________________________________________________
\section{Introduction}\label{sec:intro}
Feedback from active galactic nuclei (AGNs) is expected to have a significant impact on the formation and evolution of massive galaxies \citep[e.g.][]{SilkRees98,Harrison+17}.
AGN activity can effectively inhibit star formation (the so-called `negative' feedback) both via the injection of thermal energy into the circumgalactic medium, which delays its cooling and subsequent accretion onto the host \citep{CiottiOstriker97,Fabian+12}, and via the production of galaxy-scale outflows capable of driving a significant fraction of the gas reservoir out of the galaxy \citep{Zubovas+12,Pontzen+17}.
On the other hand, AGNs can also promote star formation (`positive' feedback) by enhancing gas pressure in the interstellar medium \citep{Silk13,Cresci+15}.

Virtually all current theoretical models of galaxy evolution include AGN feedback as a key ingredient to explain the properties of high-mass galaxies at various redshifts \citep{King03,BoothSchaye09,Weinberger+17}, and current galaxy-scale hydrodynamical simulations predict a multi-phase medium around AGN-hosting systems with properties compatible to those observed \citep[e.g.][]{Gaspari+17,Ciotti+17,Richings+18,Gaspari+20}.
However, despite the widespread success of these models, the physics of the interplay between AGN and the interstellar medium (ISM) remains elusive, with different possible candidates for both the initial feedback mechanism (winds, jets, radiation) and the mode by which feedback communicates with the gas reservoir \citep[energy or momentum-driven feedback, for a comprehensive review see][]{KingPounds15}.
%Dedicated multi-wavelength observations and modelling of the wind properties at different physical scales are therefore required to better understand the co-evolution between AGNs and their hosting galaxies.

From an observational point of view, AGN-driven outflows are commonly detected virtually at all redshifts using a variety of tracers covering the entire electromagnetic spectrum \citep{Cicone+18}.
Optical emission lines such as \ha\ and \oiii$\lambda5007$ have been routinely used to study the warm ($T\sim10^4\K$) ionised component of the wind in tens of thousands low-$z$ AGNs \citep[e.g.][]{Woo+16} and in a few hundreds of high-$z$ systems, up to $z\simeq3.5$ \citep{Harrison+12,Cresci+15,Brusa+15,Carniani+15,Bischetti+17}.
Deep sub-mm and radio observations have revealed the presence of a neutral (both atomic and molecular) component of the wind as traced mainly by \hi, CO and OH lines \citep{Morganti+05,Feruglio+10,Maiolino+12,Cicone+14,Gonzalez-Alfonso+17,Fluetsch+19}, which is thought to contribute substantially to the overall mass outflow budget especially in low-luminosity AGNs \citep{Fiore+17}.
Outflows of neutral atomic gas can also be seen at optical wavelengths via the NaID line \citep[e.g.][]{Baron+20}.
However, while fundamental to characterise the properties of the wind on galaxy-scale, these data carry little information on how such winds generate from within the active nucleus and propagate into the ISM.

X-ray data instead allow to dig deeper within the nuclear region and provide fundamental clues on the wind properties at sub-pc scales via the study of blue-shifted Fe K high-ionisation absorption lines \citep{Pounds+03,Reeves+09,Gofford+13,Nardini+15,Gofford+15}.
In particular, a comprehensive study of a sample of 42 Seyferts observed with XMM-Newton carried out by \citet{Tombesi+10b,Tombesi+11,Tombesi+12,Tombesi+13} reveled that about $40\%$ of the sources host so-called Ultra Fast Outflows (UFOs). These are X-ray winds with mild-relativistic velocities ($v_{\rm out}\sim0.1c$) and very large kinetic luminosity ($\sim10^{42}-10^{45}\ergs$), potentially enough to quench star formation within their host \citep{HopkinsElvis10}.
Clearly, targeting galaxies hosting UFOs with multi-wavelength observations is an excellent strategy to study the AGN-ISM interplay on different physical scales. Connecting the properties of nuclear X-ray winds with those of the galactic-scale outflows is critical to understand their acceleration and propagation mechanisms and, ultimately, their effect on the host galaxy.

The goal of this study is to relate the properties of galaxy-scale ionised outflows, traced by optical emission lines, to those of X-ray UFOs, in two nearby ($z\!=\!0.06$) bright QSOs, MR\,2251-178 and PG\,1126-041.
In particular, we take advantage of the unprecedented resolution, sensitivity and spectral coverage offered by MUSE in its adaptive optics (AO)-assisted narrow field mode (NFM) to map the outflow components and to compute in detail their energetics.
%We show that momentum-driven wind propagation is the favoured mechanisms relating feedback from sub-pc to galaxy scales.

This paper is structured as follows.
In Section \ref{sec:data} we present our observations and data reduction strategies.
In Section \ref{sec:method} we describe the approach adopted to derive the properties of the ionised gas from the MUSE data.
Our results on the kinematics and energetics of the ionised gas are presented in Section \ref{sec:results}, with further discussion in Section \ref{sec:discussion}.
Conclusions and final remarks are drawn in Section \ref{sec:conclusions}.
Throughout this work we adopt a $\Lambda$CDM flat cosmology with $\Omega_{\rm m,0}=0.27$, $\Omega_{\Lambda{\rm,0}}=0.73$ and $H_{\rm 0}=70\kmsMpc$.

\section{Data} \label{sec:data}
\subsection{Observed targets}
Our sample consists of two QSOs hosting the most powerful, well-established X-ray winds at $z\!<\!0.1$ that were visible in the southern sky at the time of the MUSE observations.
Details on these objects are provided below.

\subsubsection*{MR\,2251-178}
This system is a radio-quiet, type-1 luminous ($\log(L_{\rm BOL}/\lsun)\!\sim\!12.15$, see Table \ref{tab:properties}) quasar, the first QSO to be identified via X-ray observations \citep{Cooke+78,Ricker+78}, and the first system for which X-ray absorption by warm photoionised gas was established \citep{Halpern+84}.
The absorbing gas, variable in terms of ionisation state and column density \citep{Pan+90} and visible also in UV \citep{Ganguly+01}, was found to feature a high-velocity and highly-ionised outflowing component traced by \fexxvi\ \citep{Gibson+05}.
In the last decade, the sub-pc scale UFO of MR\,2251-178 has been characterised in detail thanks to several X-ray observations at different resolutions, energies and epochs, indicating spatially and temporally variable outflows with velocities in the range $0.04-0.14c$ \citep{Gofford+11, Gofford+13, Reeves+13}.

The circumgalactic environment of MR\,2251-178 is peculiar. 
The QSO lies in the outskirt of a loose, irregular cluster \citep{Phillips80} and is surrounded by a giant ($200\kpc$-wide) emission-line nebula detected in \oiii\ and \ha\ \citep{Bergeron+83,diSeregoAlighieri+84,Macchetto+90}.
The nebula is characterised by a regular kinematics \citep{Shopbell+99} but in apparent counter-rotation with respect to the gas located within the host galaxy \citep{Norgaard-Nielsen+86}.
While the filamentary morphology of the nebula resembles that of the gaseous envelopes surrounding powerful radio galaxies, this QSO shows only a weak radio emission with a double jet-like appearance \citep{Macchetto+90}.
Both internal mechanisms (like AGN-driven winds) and external ones (merger debris, cooling flow) have been proposed as possible origins for this extended gaseous structure.
We discuss further the nature of this nebula in Section \ref{ssec:nebula}.

\subsubsection*{PG\,1126-041}
This object, also known as Mrk\,1298, is a radio-quiet AGN, with a luminosity in between those of typical Seyferts and QSOs ($\log(L_{\rm BOL}/\lsun)\!\sim\!11.62$).
As opposed to the previous system, in PG\,1126-041 the outflow component was firstly detected in UV absorption using data from the International Ultraviolet Explorer (IUE) by \citet{Wang+99}, who reported outflow speeds in the range $1000-5000\kms$, while only weak features were visible in the \emph{ROSAT} X-ray spectrum. 
Later on, this QSO was the target of a multi-epoch observational campaign aimed at constraining the properties of its central engine with \emph{XMM-Newton} \citep{Giustini+11}.
These observations revealed a complex X-ray spectral variability \citep[which was previously noticed by][]{Komossa+00} on time scales of both months and hours, and the presence of a highly ionised X-ray absorber at sub-pc scales with outflow velocity of about $0.055c$, variable on time scales of a few kiloseconds.

\subsection{MUSE observations and data reduction}
MUSE data for these two objects were collected on 8 May 2019, 4 Aug. 2019 and 6 Aug. 2019, under program 0103.B-0762 (PI G.\,Cresci).
The data consisted of two Observing Blocks (OBs) for each target, for a total of eight exposures of $480$s (MR\,2251-178) and $12$ exposures of $460$s (PG\,1126-041) each, together with four $100$s sky exposures for each galaxy. The nominal seeing derived from the DIMM measurements during the observations were $\sim0.84\arcsec$ for MR\,2251-178 and $\sim0.48\arcsec$ for PG\,1126-041.
The exposures were dithered and rotated by $90$ degrees in order to remove the pattern produced by the 24 channels associated to each IFU, as well as to optimise the cosmic rays removal and background subtraction. The sky exposures were used during the data reduction to create a model of the sky lines and sky continuum, to be subtracted from the science exposures. 
The data reduction has been carried out with the MUSE pipeline v1.6, using ESO Reflex, which gives a graphical and automated way to execute with EsoRex the Common Pipeline Library (CPL) reduction recipes, within the Kepler workflow engine \citep{Freudling+13}.

Data for MR\,2251-178 (PG\,1126-041) consist of a cube of $381\times354$ ($362\times357$) spaxels, for a total of about 135 (129) thousands spectra covering a field of view of $9.65\arcsec\times8.97\arcsec$ ($9.17\arcsec\times9.05\arcsec$).
Spaxel size is of $25\times25\,{\rm mas}^2$.
Each spectrum is sampled by $3681$ channels covering the spectral range $\lala4750-9350\AA$ with a channel width of $1.25\AA$.
Spectral resolution in this range varies almost linearly from $R\sim1750$ (FWHM of $\sim170\kms$) in the blue to $R\sim3600$ (FWHM of $83\kms$) in the red.
Details on the MUSE spatial resolution are given in Table \ref{tab:resolution} and discussed in Appendix \ref{app:MUSE_PSF}.

\subsection{General properties of the QSOs}
Table \ref{tab:properties} lists the main physical properties of the two QSOs, most of which are derived in the present study.
The new AO-assisted data provide a more refined estimate for the coordinates of the two QSOs, for which we take the peak intensity observed at rest-frame $\lambda5100\AA$, and for their redshift, for which we measure $z=0.064$ in MR\,2251-178 and $0.060$ in PG\,1126-041 using the bright \oiii\ doublet towards the galactic nuclei.
We note that, at these $z$, the MUSE NFM resolution of $\sim0.1\arcsec$ (FWHM) corresponds to a spatial scale of about $120\pc$, which is perfectly adequate to describe the detailed structure and kinematics of the ionised gas in the central $12\times12\kpc^2$, corresponding to the MUSE field of view.
%We discuss in more depth the MUSE point spread function (PSF) in Section \ref{ssec:blr}.

The quoted $L_{\rm BOL}$ for MR\,2251-178 is the mean of two distinct measurements, the first using the rest-frame $\lambda5100\AA$ luminosity with bolometric correction from \citet{Richards+06}, the second from the $2-10$ keV luminosity \citep{Nardini+14} using the bolometric correction from \citet{Lusso+10}, which depends on the optical-to-X-ray spectral index, $\alpha_{\rm ox}=-1.05$.
Similarly, in PG\,1126-041 we take the mean between the $L_{\rm BOL}$ derived from the $2-10$ keV luminosity \citep{Giustini+11} using $\alpha_{\rm ox}=-1.62$, and that derived by \citet{Veilleux+13} using the combined $\lambda5100\AA$ and $8\!-\!1000\,\mu{\rm m}$ luminosities.
The uncertainties on $\log(L_{\rm BOL})$ reported in Table \ref{tab:properties} are given by half the difference between the two measurements.

Following \citet{VP06}, black hole masses $M_{\rm BH}$ are determined from our MUSE data using the FHWM of the component of the \hb\ line coming from the broad line region (BLR, see Section \ref{ssec:nuclear}), and either the BLR \hb\ luminosity or the rest-frame $\lambda5100\AA$ luminosity.
The two methods give perfectly compatible results, and we take the scatter in the relations of \citet{VP06} as the main uncertainty on these values.

The velocity $v_{\rm UFO}$ and column density $N_{\rm H,UFO}$ of the gas associated to the sub-pc scale UFOs are taken from previous X-ray studies.
For PG\,1126-041 we adopt $v_{\rm UFO}=0.055c$ and $N_{\rm H,UFO}=7.5\times10^{23}\cmmq$ from \citet{Giustini+11}.
In MR\,2251-178, at least two distinct high-speed components are detected in X-ray, the first with $v_{\rm UFO}=0.14c$ and $N_{\rm H,UFO}=5\times10^{21}\cmmq$ \citep{Gofford+11}, the second with $v_{\rm UFO}\sim0.052c$ and $N_{\rm H,UFO}>1.5\times10^{23}\cmmq$ \citep{Reeves+13}, where only a lower limit on $N_{\rm H,UFO}$ can be determined given the degeneracy with the ionization parameter in the line modelling.

\begin{table}
 \centering
% \begin{minipage}{140mm}
   \caption{Main physical properties of the two QSOs presented in this work.}
   \label{tab:properties}
   \begin{tabular}{lcc}
   \hline
   \hline
                                & MR\,2251-178      & PG\,1126-041 \\
     \hline
     R.A.\,(J2000)              & $22^{\rm h}54^{\rm m}05.936^{\rm s}$ & $11^{\rm h}29^{\rm m}16.704^{\rm s}$ \\
     Dec.\,(J2000)              & $-17^{\circ}34'53.87\arcsec$ & $-04^{\circ}24'06.66\arcsec$ \\
     $z$                        & $0.064$      & $0.060$\\
     $D_{\rm L}/\Mpc$           & $287.8$       & $269.0$ \\
     scale                      & $1.23\kpc/\arcsec$        & $1.16\kpc/\arcsec$ \\
     $\log(L_{\rm BOL}/\lsun)$  & $12.15\pm0.11$  &  $11.62\pm0.15$ \\
     $\log(M_{\rm BH}/\msun)$   & $8.43\pm0.43$   & $7.75\pm0.43$ \\
     $v_{\rm UFO}/c$            & $0.14,\,0.052$   & $0.055$ \\
     $N_{\rm H,UFO}/10^{23}\cmmq$   & $0.05,\,\gtrsim1.5$ & $7.5$ \\
     \hline
   \end{tabular}
% \end{minipage}
\end{table}

\section{Method} \label{sec:method}
We now describe in detail the approach adopted to derive the properties of the ionised gas from the sky-subtracted flux calibrated MUSE data cubes.
Our method consists of three steps.
In the first step we focus on modelling the broad emission lines originating from within the nuclear regions of our galaxies (the so-called `broad-line regions', BLR).
In the second step, the model previously built is used as a template to remove the contamination of the (unresolved) nuclear emission from all the spaxels in the MUSE data, caused by the extended wings of the PSF. 
Finally, in the last step we focus solely on the (narrow) emission lines from the diffuse ionised gas.

\subsection{Modelling the BLR spectrum}\label{ssec:nuclear}
The goal of this step is to extract a template model for the Balmer (\ha\ and \hb) and \feii\ emission lines originating from within the BLR.
This step is crucial as our results are dependent on how well the fainter emission from diffuse gas can be separated from the brighter BLR component.
Hence, it is imperative to select and model a portion of the data domain where emission lines from both BLR and diffuse gas have a high signal-to-noise ratio (S/N) and can be visually identified and separated.
The central spectrum, which we take as the spectrum integrated within a radius of 2 spaxels from the system centre, can be used for this purpose in MR\,2251-178. 
In PG\,1126-041 we take instead the spectrum integrated within 2 spaxels at a small offset ($+0.2\arcsec$ in DEC) with respect to the centre, because at this location the narrow \nii\ doublet, originating from within the diffuse gas, is better recognisable on top of the broad \ha\ line from the BLR.
This allows us to better separate the contribution of different components (\ha\ from the BLR and from the diffuse gas, \nii\ from the diffuse gas) that are severely blended in the central spectrum.
For simplicity, in what follows we refer to these spectra as `nuclear', in spite of the differences described.

We model the nuclear spectrum via a hybrid approach that combines the use of analytical functions, spectral templates and ad-hoc adjustments driven by the data.
The spectrum is decomposed into the sum of different components: continuum emission from stars and AGN, broad ($\gtrsim1000\kms$) Balmer and \feii\ emission lines originating from the BLR, and narrow ($\lesssim1000\kms$) emission lines from different atomic species (\oiii, \ha, \hb, \nii, \sii\ and \oi).
All these components are fit to the data simultaneously in the wavelength range $\lala4700-7000\,\AA$ using the following assumptions.

As in our systems the AGN outshines the light from stars, we do not model the stellar continuum in detail but use instead a third-order polynomial to approximate the overall continuum underneath the emission lines.
A third-order polynomial allows us to reproduce the large-scale modulations visible in our spectra without affecting the fit of the individual lines.

Following \citet{Cresci+15}, we describe the \ha\ and \hb\ BLR lines as broken power-law distributions convolved with Gaussian functions.
These particular functional forms allow us to model asymmetric line profiles with a variable degree of `cuspness' at their centre.
Given that the broad \ha\ and \hb\ lines show clear differences in our nuclear spectra, we choose to model each line independently.
\feii\ BLR lines are modelled using the semi-analytic templates of \citet{Kovacevic+10}, which comprise a total of $65$ individual emission lines at $\lala4000-5500\,\AA$.
These lines are grouped within five distinct sets (four of which are fully theoretical, while the fifth is based on the spectrum of I Zw $1$), whose relative intensity is free to vary while we impose velocity shift and width to be the same for all lines.

We describe the diffuse gas emission from individual species using multiple Gaussian components, each having its own centroid, amplitude and width.
Individual line profiles can be complex in this sources, for instance a minimum of three components are required to reproduce the shape of the \oiii\ line.
Once decided the number of components $N$ required to reproduce a given `reference' line, we make the strong assumption that all the various atomic species trace the same emitting region, thus imposing that the velocity-profile of each individual line is a rescaled version of the reference one\footnote{Lines are re-scaled in the velocity space, not in the wavelength space}.
For both QSOs examined, the adopted reference line is the bright \oiii$\lambda5007$ emission line, which does not suffer from significant contamination from the BLR.
We neglect variations in the shape of the different lines induced by the MUSE variable spectral resolution, since typical line-widths measured in the nuclear regions are always larger (by a factor $\gtrsim3$) than the instrumental broadening.
This significantly decreases the number of degrees of freedom for the diffuse gas modelling down to $3N+m-1$, with $m$ being the number of atomic species considered.
This stratagem is crucial to break the degeneracy between BLR and diffuse emission and to provide physically meaningful results (see Section \ref{sec:discussion}).
We note that this assumption is used only in this particular step, and we drop it when modelling the diffuse gas alone (Section \ref{ssec:lines}).

We set-up our model and fit it to the data using our own \texttt{python} routines\footnote{Specifically we make use of the \texttt{curve\_fit} routines from the \texttt{scipy} library.}.
We make then a final refinement.
We assume that, in the regions dominated by the \ha\ and \hb\ lines (approximately at $\lala4730-4930\AA$ and $\lala6400-6680\AA$ rest-frame), deviations from our best-fit model are solely due to limitations in our BLR description.
This assumption is justified by the fact that the broad wings of the Balmer lines show quite a few `bumps' (especially in MR\,2251-178), indicating that the assumed broken power law description does not fully capture their complexity. 
Under this assumption, we use the residual from our best-fit profile as a  correction term to our BLR model in the region of the Balmer lines, so that our final model matches perfectly the data in those regions.
The use of this data-driven correction term implies the injection of a small amount of noise into our model.
However, among all spaxels in the datacube, our nuclear spectra have the highest possible S/N, thus their use as a template to clean the data from BLR emission does not bias our analysis.
%Also, we stress that this refinement gives only a minor correction to the overall shape of model profile (see below).
Clearly, this BLR correction term may contain important information on the physics of the BLR, whose study is however beyond the purpose of this work.

\begin{figure*}
\begin{center}
\includegraphics[width=0.90\textwidth]{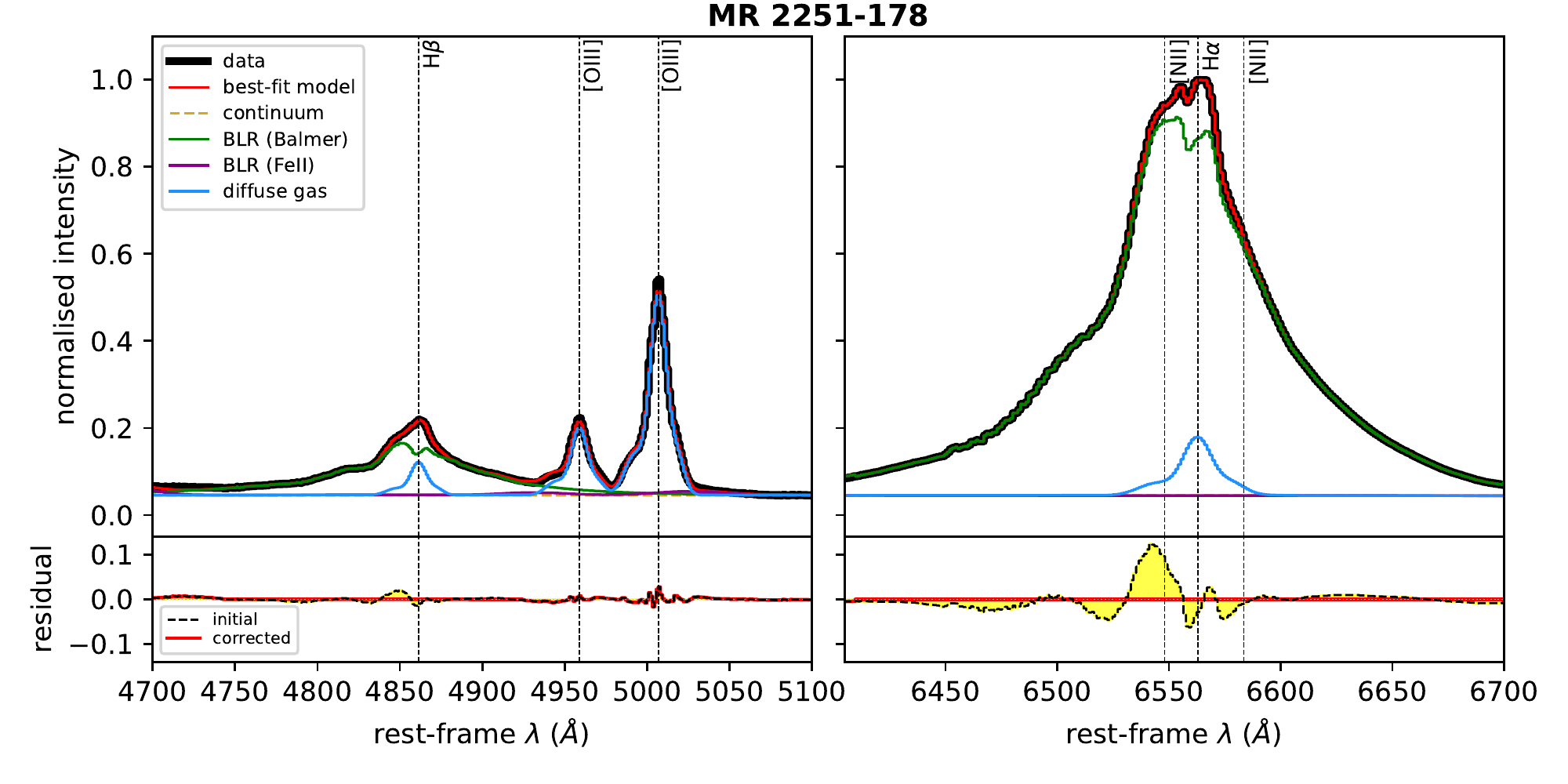}
\includegraphics[width=0.90\textwidth]{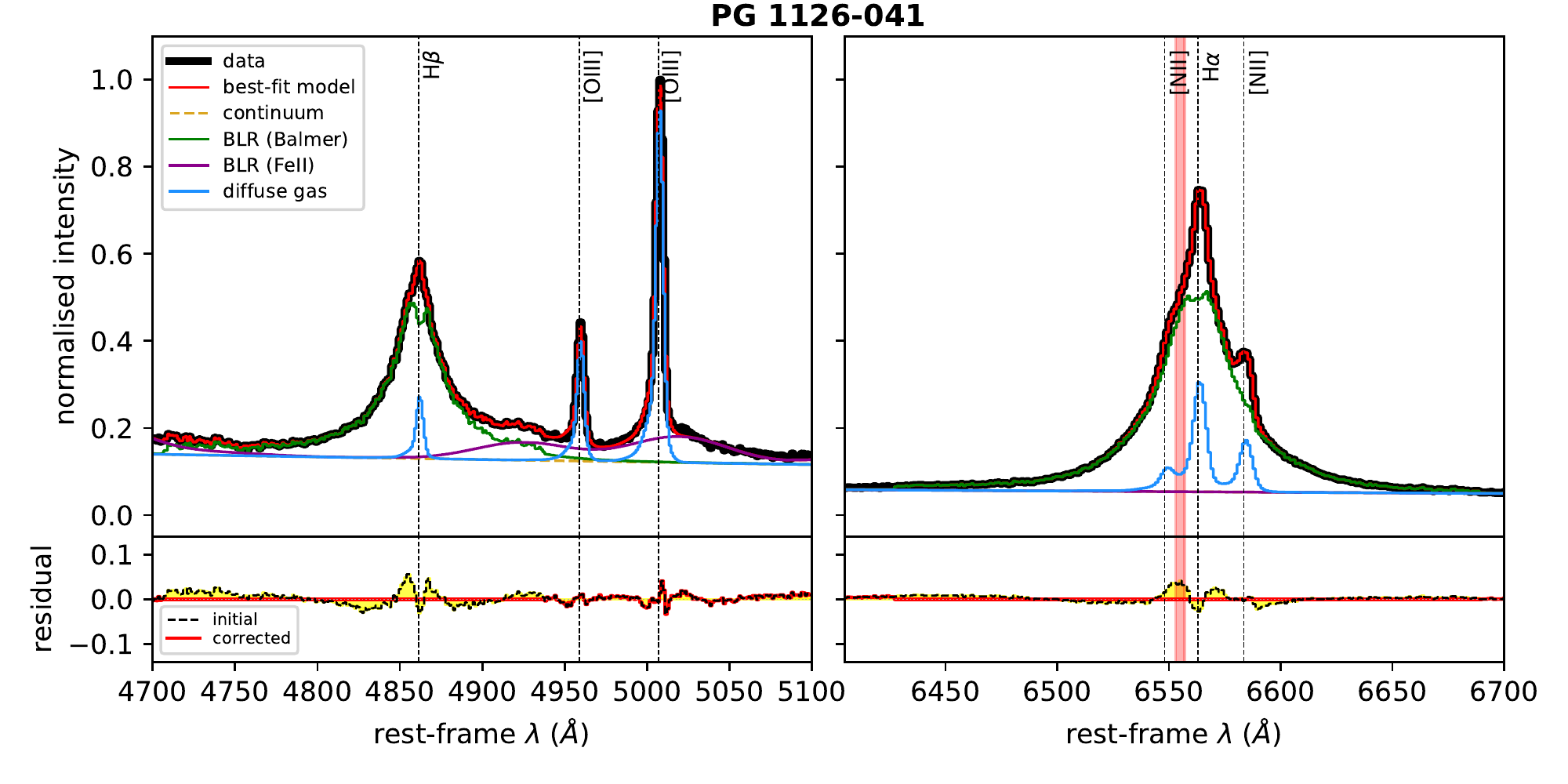}
\caption{Representative regions of the optical nuclear spectrum (solid black lines) for MR\,2251-178 (\emph{upper panels}) and PG\,1126-041 (\emph{lower panels}), along with their respective best-fit models (solid red lines). The left-hand panels focus on the \hb\ and \oiii\ lines, while the right-hand panels are centred on the \ha\ and \nii\ lines.
The models are given by the sum of the \feii\ and Balmer emission lines from the nuclear region (purple and green solid lines respectively), diffuse gas emission from individual atomic species (blue solid lines) and a third-order polynomial (dashed yellow lines). Emission from diffuse gas is modelled with three Gaussian components.
Residuals (data-model) are shown below each panel; red solid/black dashed lines are used for models with/without the BLR correction term (see text).
The red vertical band in the bottom-right panels marks a region contaminated by sky-subtraction residual and is not fit by our model.} 
\label{fig:nuclear}
\end{center}
\end{figure*}

We show the results of our modelling in Figure \ref{fig:nuclear}, where we focus on two regions encompassing the brightest lines (\hb\ and \oiii\ in the left panels, \ha\ and \nii\ in the right panels) in the nuclear spectra of MR\,2251-178 (top panels) and PG\,1126-041 (bottom panels).
The models (red solid lines) reproduce the data (black solid lines) remarkably well, also thanks to the (small) correction term added to the BLR emission.
The magnitude of this term can be seen in the inset below each panel, which zooms-in on the residuals (data-model) derived with or without such correction.
In general, the amplitude of the correction is of the order of the residuals in the \oiii\ line, the exception being the \ha\ line in MR\,2251-178 whose profile shape is very complex and where the correction gets larger.
Interestingly, after this data-driven adjustment, the Balmer BLR lines appear to have a double-peaked shape, reminiscent of that produced by an unresolved rotating disc.
By construction, all emission lines from diffuse gas are a re-scaled version of the \oiii$\lambda5007$ line.
Blue-shifted wings can be seen in the \oiii\ lines, indicating the presence of an outflow.
In PG\,1126-041, \feii\ emission fully accounts for the extended red wing close to the \oiii\ line.

\subsection{Removing the contribution from the BLR}\label{ssec:blr}
We now proceed to remove the contamination of the BLR emission from the entire dataset.
As the BLR is unresolved, we expect its emission to be spatially distributed as the MUSE PSF.
The latter, given the AO-nature of our data, should be roughly made by a combination of an AO-corrected narrow core surrounded by a broad (seeing-limited) halo.
The presence of this halo is particularly relevant as bright sources - such as the BLR - can contaminate spaxels located far beyond the nominal spatial resolution.
The exact shape of the PSF strongly depends on the observing conditions and is difficult to determine a priori, especially for AO observations.

To address this problem, we use the software \ppxf\ \citep{Cappellari17} to model the whole datacube under the assumption that the contribution of the BLR emission to each spectrum is only a re-scaled version of that determined within the nuclear region.
In practice, for each system we produce a BLR template made by the sum of the best-fit \feii\ lines, Balmer lines and the polynomial found in Section \ref{ssec:nuclear}.
We then use \ppxf\ to fit each spectrum in the datacube with a combination of our BLR template, which can be only re-scaled in amplitude, and multiple Gaussian components for the spatially resolved diffuse emission.
With respect to the previous step, we allow for a larger freedom in the modelling of the diffuse gas by assuming that the flux ratios of the different Gaussian components can vary freely amongst the different atomic species, so that different lines can have different total profiles. 
We still constrain the central velocities and velocity widths of all components to be the same in all the species considered, under the assumption of a unique underlying large-scale kinematics.
Finally, the BLR component of the fit is subtracted to the original datacube, producing a `cleaned' cube containing only narrow lines from the various atomic species.

\begin{figure}
\begin{center}
\includegraphics[width=0.5\textwidth]{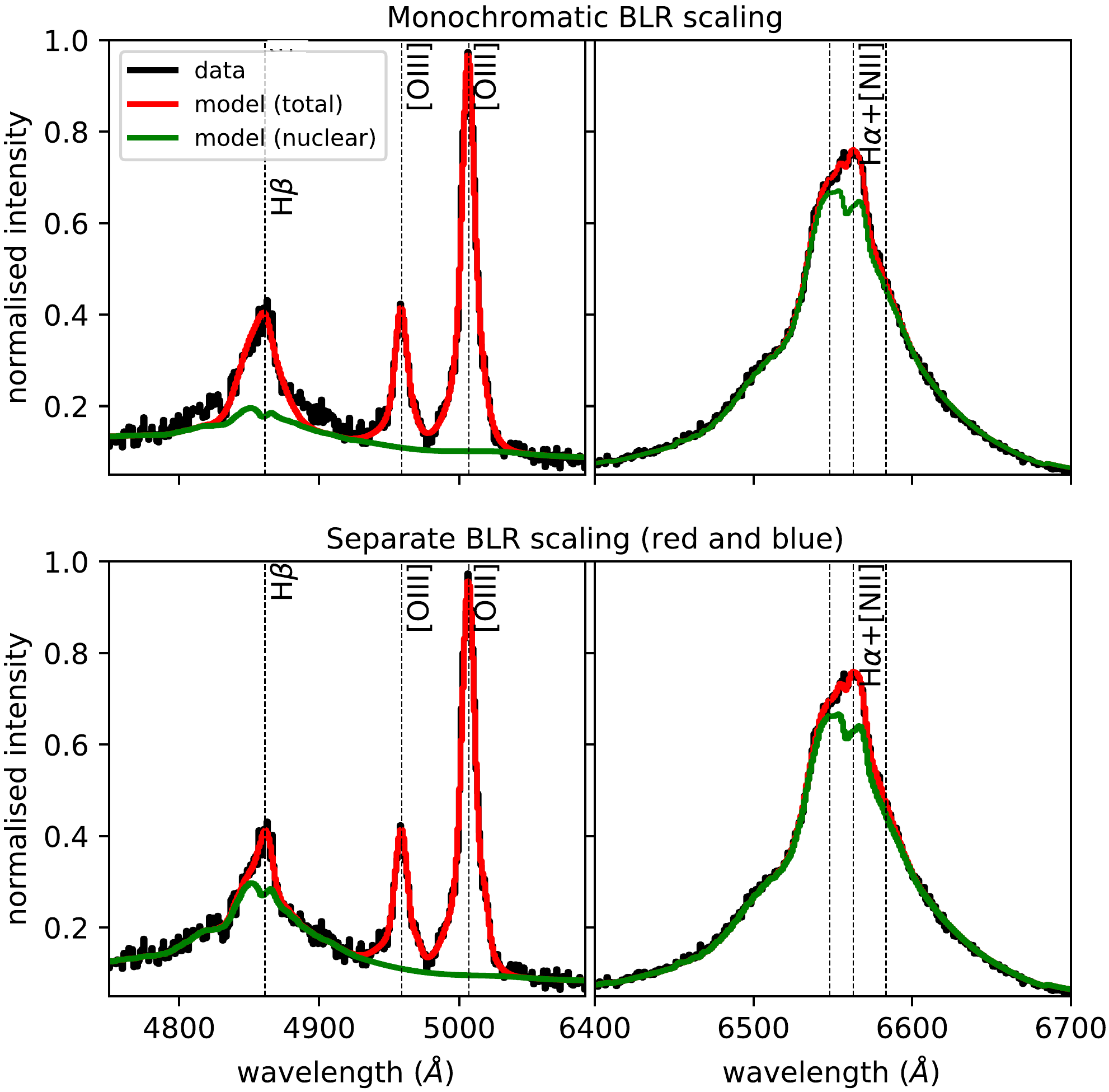}
\caption{Representative spectrum taken at $0.34\arcsec$ from the centre of MR\,2251-178 (black lines in all panels), and best-fit models (red lines) derived by re-normalising the BLR contribution either as a whole (top panels) or for the red and blue regimes separately (bottom panels), as discussed in the text. The green lines show the contribution of the BLR emission to the full spectrum. The separate normalisation leads to a much better fit to \hb\ line (left panels).} 
\label{fig:blr_comparison}
\end{center}
\end{figure}

A couple of factors, however, complicate this procedure.
A first concern is that the described approach leads to a poor fit to the spaxels around the centre, where the broad \ha\ and \hb\ components appear to give different contributions to the spectrum with respect to what we inferred from the nuclear region.
This issue is a consequence of the variation of the PSF with wavelength due to AO-assisted nature of our data, with the Strehl ratio being higher in the red (\ha) than in the blue (\hb).
To account for this, we break the assumption of rigid re-normalisation of the full BLR model, which is instead split into two sub-regions (above and below $5700\AA$ rest-frame) that can scale separately in the fit.
Figure \ref{fig:blr_comparison} shows a representative spectrum of MR\,2251-178 to demonstrate that the approach described leads to a much better fit to the data in the regions around the \hb\ line.
Details on the MUSE PSF in the two spectral regions are provided in Appendix \ref{app:MUSE_PSF}.

A second concern is the number of narrow components required to reproduce the data.
This can vary across the field due to either an intrinsic variation in the complexity of the line profile, or to variations in the signal-to-noise ratio.
In order to decide the optimal number of components that are used in each spaxel, we adopt the following statistical approach.
We first fit the whole datacube multiple times, using each time a fixed number of narrow components ranging from one to three.
We found this range to be optimal given that even the most complex profile in our data can be accurately described by three components, while a single component is sufficient to model the fainter lines in the external part of the field of view, where the S/N is small and the dynamics of the system simpler.
Then, for each spaxel we compare the distribution of the residual (data-model) in a wavelength range encompassing the bright \oiii\ line using a Kolgomorov-Smirnov (KS) test.
In practice, we ask whether the residuals of a model with $n+1$ components differ statistically from those of a model with $n$ components, starting from $n=1$.
If they do, we set $n\!=\!n+1$ and re-do the test, otherwise we set to $n$ the optimal number of components for that particular spaxel.
This method does not rely on any particular likelihood estimator, and accounts only for the relative improvements that an additional fitted component would bring to the model.
We can change our acceptance tolerance by varying the $p$-value threshold\footnote{$0\!\leq\!p\!\leq\!1$, where $0$ ($1$) minimises (maximises) the number of components required.} of the KS test. 
We experimented with different values and verified visually their effect on the final model, finding that $0.6\!<\!p\!<\!0.9$ is the optimal range to describe our data.
The fiducial $p$-values adopted are $0.75$ for MR\,2251-178 and $0.8$ for PG\,1126-041.
Maps showing the number of components are presented in the Section below.

\subsection{Modelling the diffuse emission}\label{ssec:lines}
In this step we focus on the continuum and BLR-subtracted cubes to extract the properties of the emission lines from the diffuse ionised gas.
While the latter were already modelled by \ppxf\ in the procedure above, re-fitting these lines on the cleaned dataset brings two advantages.
The first is that we have less free parameters as we do not need to model the complex BLR emission.
The second advantage consists in a larger freedom in selecting the desired functional forms to fit the data, bypassing the limitations of \ppxf\ of using solely Gauss-Hermite functions.
In practice, however, we found that a multi-Gaussian representation of the line profiles is perfectly satisfactory for the two datasets in exam.
In order to increase the S/N in the external regions of the datacube we employ a Voronoi tessellation of our data using the method of \citet{CappellariCopin03}.
This consists in generating an adaptive spatial binning, with grid elements that can vary in size and shape, ensuring a minimum S/N ratio across the 
whole dataset.
We require that each Voronoi bin has a minimum S/N of $15$ in the \oiii\ line\footnote{computed as the integrated S/N in the rest-frame $\lala5000-5015\AA$ range.}.
The multi-Gaussian fit follows the same statistical methodology described in Section \ref{ssec:blr}.
In this step we model explicitly the following emission lines: \hb, \oiii$\lambda4959$, \oiii$\lambda5007$, \nii$\lambda6548$, \ha, \nii$\lambda6583$, \sii$\lambda6716$ and \sii$\lambda6731$.
We treat the \oiii\ and \nii\ lines as doublets, fixing both the \oiii\ $\lambda5007/\lambda4959$ line ratio and the \nii\ $\lambda6583/\lambda6548$ line ratio to the theoretical value of $3$.
Fainter lines, such as \hei, \oi$\lambda6300$ and \oi$\lambda6364$, are detected (especially around the nuclear region) but not modelled.

The line shapes can be complex as resulting from the mixture along the line-of-sight of gas that follows the galaxy's global kinematics with material showing strong `peculiar' motions (such as in an outflow).
As the main target of our study is the outflowing gas, which is commonly traced by the broader component of the lines, we need to develop an automated method to separate it from the remaining `non-peculiar' kinematic component.
For this purpose, we post-process our best-fit model of the diffuse gas as follows.

For any given line profile, we first determine its peak velocity $v_{\rm peak}$ as the velocity corresponding to the line peak.
We then flag as `high-velocity' (HV) all Gaussian components that have at least a fraction $f_{\rm HV}$ of their flux at relative velocities $|v-v_{\rm peak}|$ above a given threshold $v_{\rm HV}$.
If more than one component satisfies the condition above, these are merged together in a single HV component, and we label as `low-velocity' (LV) the remaining portion of the line profile.
This classification based on $v_{\rm peak}$ is well suited for our QSOs where emission lines from different species peak at similar velocities and the HV components are always sub-dominant, but it may not be optimal in cases where $v_{\rm peak}$ varies significantly from one line to another or when the flux is dominated by the HV components.
We note that, with the classification adopted, even lines described by a single component can be classified as HV if their velocity width is sufficiently large.
After experimenting with different values, by visual inspection of the profiles we found that $f_{\rm HV}\!=\!0.4$ and $v_{\rm HV}\!=\!300\kms$ give the optimal separation between the HV and LV components.
These values correspond to the requirement that about half of the flux in each HV component lies at velocities $\gtrsim3-4\sigma$ from the line peak, with $\sigma$ being the typical velocity dispersion of the LV component ($\sim80\!-\!100\kms$, see moment-2 maps in Fig.\,\ref{fig:kinematics}).

\begin{figure*}
\begin{center}
\includegraphics[width=0.7\textwidth]{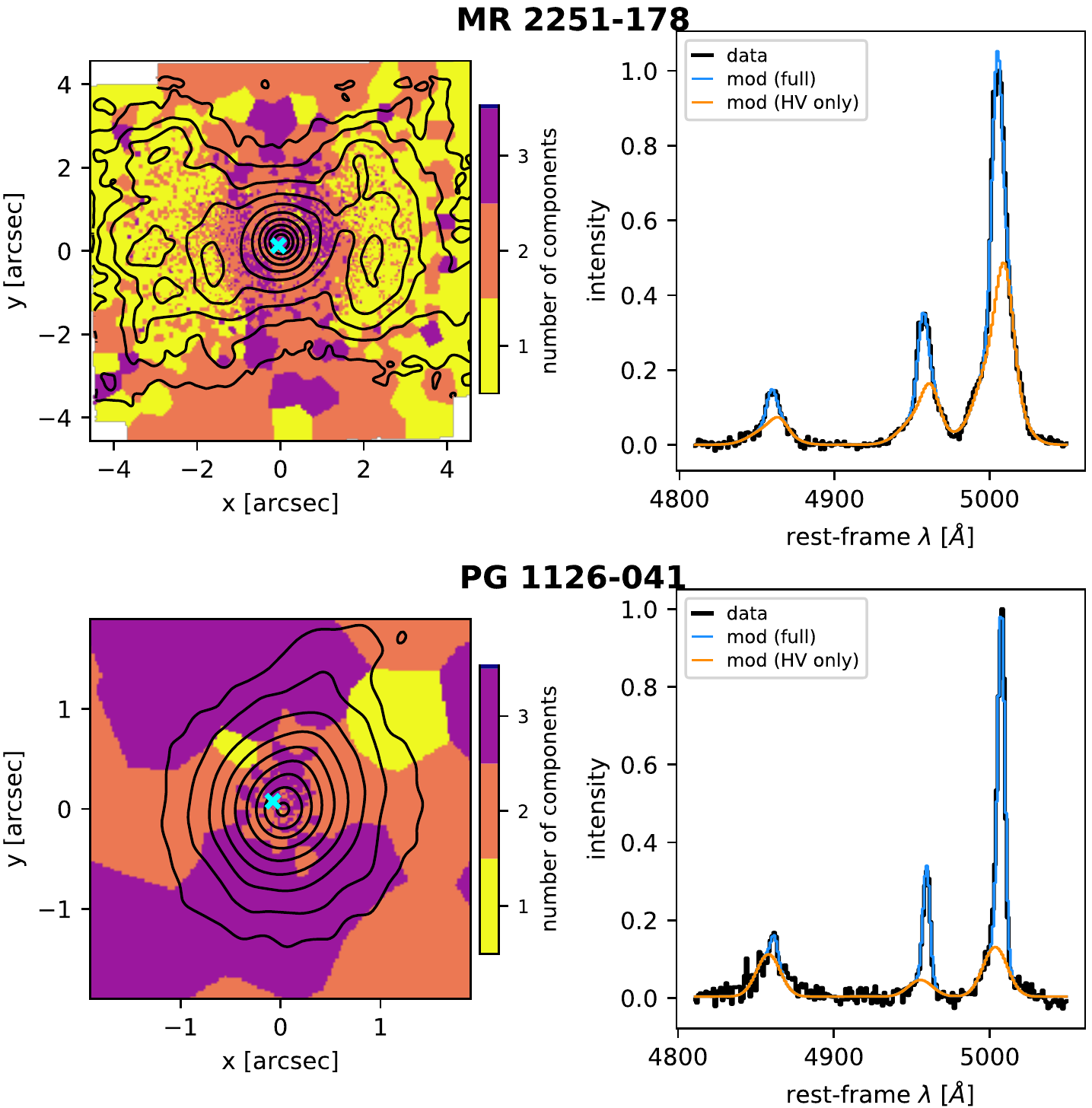}
\caption{Multi-component fit for the diffuse gas emission lines of MR\,2251-178 (top panels) and PG\,1126-041 (bottom panels). \emph{Left-hand panels}: iso-intensity contours for the integrated \oiii\ line (black lines), the background colours show the number of components used in each Voronoi bin. The \oiii\ maps are derived by integrating the (unbinned) BLR-subtracted cubes in the rest-frame range $\lala4930-5030\AA$, and are spatially smoothed to a FWHM resolution of $0.25\arcsec$. Contours are spaced by a factor of $2$, the outermost being at $3\times$ the rms-noise. The crosses in cyan mark the bins at which the spectra shown in the right panels are extracted. \emph{Right-hand panels}: representative spectra in a range encompassing the \hb\ and \oiii\ lines for the BLR-subtracted cubes (black lines), our best-fit models (blue lines) and their HV sub-components alone (orange lines).}
\label{fig:ncomp_profiles}
\end{center}
\end{figure*}

Fig.\,\ref{fig:ncomp_profiles} synthesises our modelling procedure for the diffuse gas.
The number of components used in each Voronoi bin varies across the field (left-hand panes) and typically peaks in the central regions where the S/N is higher, decreasing at larger radii.
MR\,2251-178 shows a clear dichotomy where fewer components are required to reproduce the emission in the east-west direction (where the dense shells of the nebula lie) while more components are needed in the direction perpendicular to it.
We discuss this further in Section \ref{ssec:kinematics}.
Individual line profiles are shown in the right-hand panels of Fig.\,\ref{fig:ncomp_profiles}.
These are well fit by our model, and in particular the HV component traces accurately the extended blue-wings of the emission lines.
In the Section below we use these HV components to infer the properties of the outflowing gas.

\section{Results}\label{sec:results}

\subsection{Distribution and kinematics of ionised gas}\label{ssec:kinematics}
\begin{figure*}
\begin{center}
\includegraphics[width=1.0\textwidth]{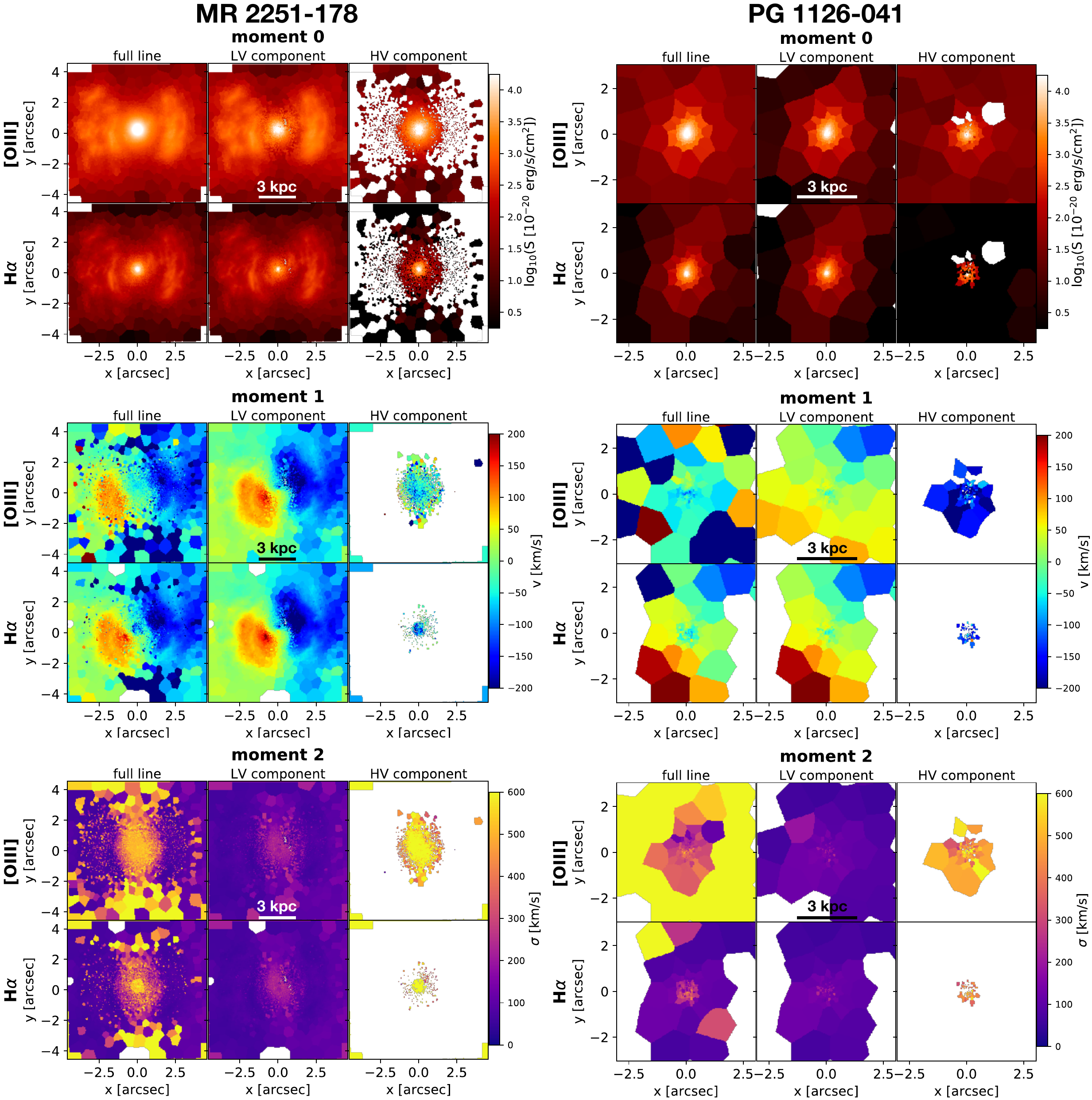}
\caption{Distribution and kinematics of the diffuse ionised gas in MR\,2251-178 (left-hand panels) and PG\,1126-041 (right-hand panels). For each system, the set of $6$ panels on top (middle, bottom) show the moment-0 (moment-1, moment-2) map for the \oiii\ (first row) and the \ha\ (second row) emission lines from diffuse gas. The different columns show the maps derived for the whole line, or for the LV and the HV components separately. Moment-1 and moment-2 maps are clipped at a peak S/N of $3$. Moment-1 maps are relative to the the systemic velocity of the QSOs, given by their estimated redshift.}
\label{fig:kinematics}
\end{center}
\end{figure*}
Fig.\,\ref{fig:kinematics} offers an overview of the distribution and the kinematics of the ionised gas in MR\,2251-178 and PG\,1126-041, as derived from our modelling of the \oiii\ and \ha\ lines.
The moment-0 (intensity), moment-1 (velocity field) and moment-2 (dispersion field) maps for the LV and HV components of these lines are shown separately, in order to better trace their distinct spatial and velocity distribution.
The moment-1 and -2 maps are clipped to a peak S/N\footnote{computed as the ratio between the peak of the line in the model (or the peak of each component, for the LV and HV maps) and the standard deviation of the residuals (data-model) within a small spectral window around the line} of $3$.
This improves significantly the quality of such maps especially close to the borders, where contamination from HV components with a low S/N is important. 
Moment-2 maps have not been deconvolved for instrumental broadening, which gives minimum observable velocity dispersions $\sigma_{\rm MUSE}$ of about $50\kms$ around the \ha\ line ansd $65\kms$ around the \oiii\ doublet.

We focus first on the LV component in MR\,2251-178, which dominates the diffuse gas emission beyond $\sim0.5\arcsec$ from the centre and traces very well the extended nebula around the galaxy.
The nebula seems to be composed by a series of thin shells of ionised gas, which wrap around the system and extend in the east-west direction up to the edge of our field of view.
Each shell shows several sub-structures which are particularly evident in the \ha\ intensity map, possibly because of the better spatial resolution achieved by the AO system at these wavelengths.
The velocity field of the nebula is peculiar: while a clear velocity gradient is visible, its direction does not match perfectly the orientation of the shells.
This mismatch suggests that the nebula cannot be described simply by a series of layers in spherical expansion from a central point, but that a more complex kinematics is at work here.
Furthermore, the moment-2 maps reveal that the material in the shell is described by an exceptionally low velocity dispersion ($\sim80\kms$), of the order of the MUSE spectral resolution, at odds with expectations from a more turbulent AGN-driven wind origin.
We discuss the origin of this nebula further in Section \ref{ssec:nebula}.
%Interestingly, the velocity field of the LV component alone is much cleaner than that of the lines taken as a whole, an indication that our low and high velocity separation method provides physically meaningful results.

The LV component in PG\,1126-041 is somewhat less spectacular.
In this galaxy the emission from diffuse gas is more centrally concentrated, and even for the brightest lines the S/N drops dramatically (and the Voronoi grid becomes very coarse) beyond the central $\sim1\arcsec$.
Still, a large-scale velocity gradient seems to be visible especially in the \ha\ line, which shows a velocity difference of about $350\kms$ across its extent.
The LV \ha\ morphology and the orientation of its velocity gradient match those of the broad-band optical image of PG\,1126-041 from the DSS\footnote{Retrievable from the NASA/IPAC Extragalactic Database, \url{https://ned.ipac.caltech.edu/}.}.
We therefore interpret this velocity gradient as that of a regularly rotating disc, whose kinematics appear to be relatively undisturbed by the presence of an active galactic nucleus.
 
While the LV components trace preferentially the large-scale kinematics, the HV components are more informative of the nuclear activity.
As expected, in both systems the HV component appears to be concentrated preferentially towards the galactic centre, featuring a negative (i.e., blue shifted) bulk velocity ranging from a few tens to a few hundreds $\kms$, and a velocity dispersion of $400-800\kms$.
These are all clear signs for the presence of AGN-driven outflows from the central regions of MR\,2251-178 and PG\,1126-041.
We note that, in both QSOs, the HV component appears to be more spatially extended in \oiii\ than in \ha. 
This is likely to be caused by the different brightness of the two lines, with the HV-component of the latter falling below our S/N threshold beyond the central $\sim1\arcsec$.
In MR\,2251-178, the outflow appears to be extended towards the north-south direction. 
The inferior S/N achieved in PG\,1126-041 prevents us from studying the detailed spatial distribution of the outflow; yet, its size appears to be at least of $\sim1\arcsec$, thus much larger than the MUSE angular resolution (Fig.\,\ref{fig:PSF_MUSE}).
The energetics of these kpc scale outflows and their connection to the sub-pc scale X-ray wind are discussed in Section \ref{ssec:energetics}.

We have focused here on the two brightest lines in our dataset, the \oiii\ and the \ha.
We have verified that the distribution and kinematics of the other atomic species that we model (\hb, \nii, \sii) are consistent with the pattern traced by these two lines, although their lower S/N strongly limits the study of the HV component.

\subsection{Outflow electron densities}\label{ssec:ne}
\begin{figure}
\begin{center}
\includegraphics[width=0.4\textwidth]{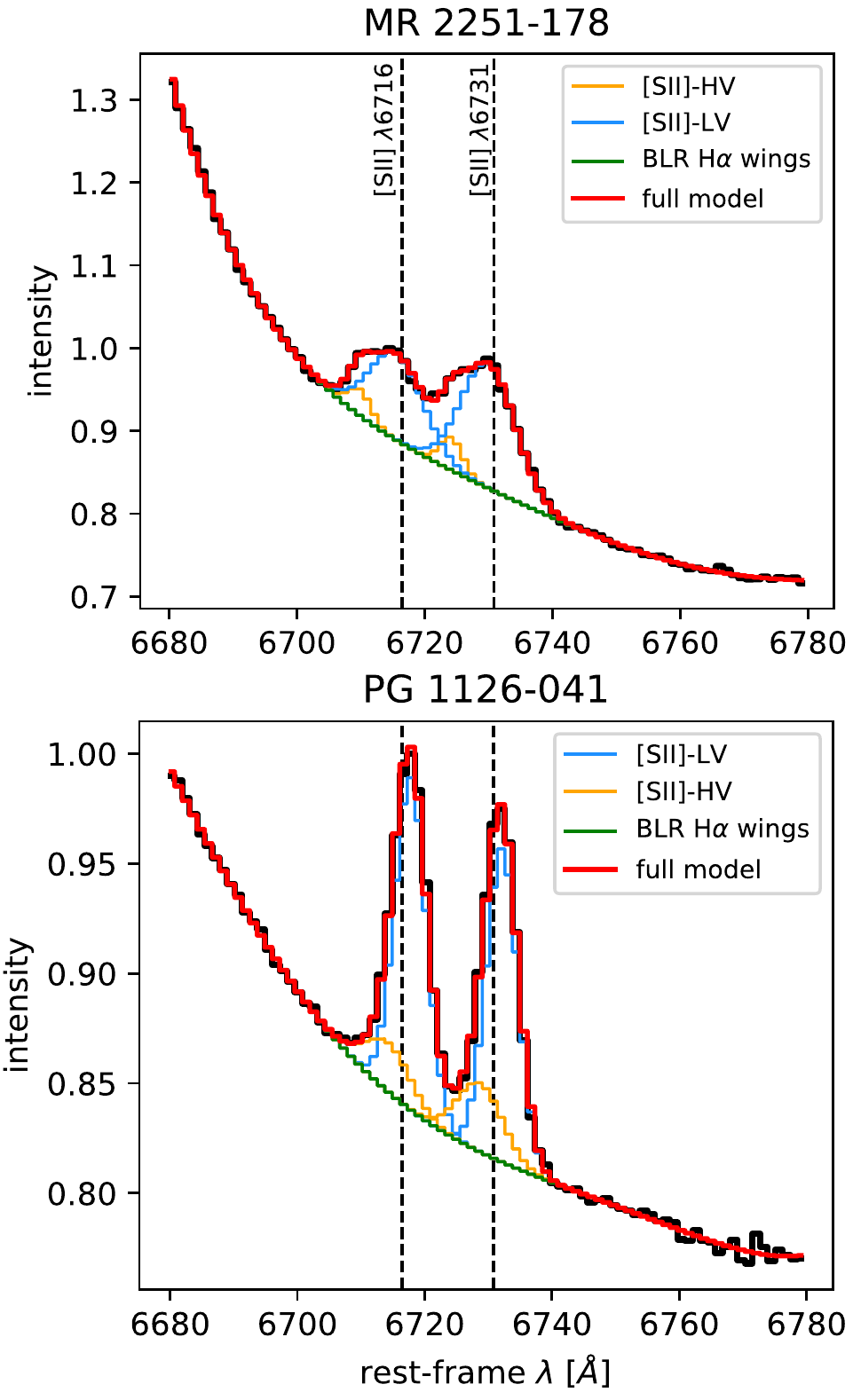}
\caption{\sii\ emission lines and related models for MR\,2251-178 (top panel) and PG\,1126-041 (bottom panel). The black lines show the spectrum extracted within an aperture of $0.05\arcsec$ (for MR\,2251-178) and $0.5\arcsec$ (for PG\,1126-041) from the systems' nuclei. The red lines show our best-fit models, with separate contributions from the wings of the BLR \ha\ (green lines), and the HV- and LV-components of the \sii\ doublets (orange and blue lines, respectively).}
\label{fig:SII}
\end{center}
\end{figure}

Converting the luminosity of the HV components into an outflow mass requires measurements of the wind electron density, $n_{\rm e}$.
In general, $n_{\rm e}$ can be estimated from the \sii$\lambda6716$/\sii$\lambda6731$ ratio \citep[e.g.][]{OF06} measured in the regions of the outflow, using the HV-components of the \sii\ lines alone.
Unfortunately, in both galaxies the \sii\ doublet sits on top of the extended (red) wing of the BLR \ha\ line, and by visually inspecting our nuclear fits (Section \ref{ssec:nuclear}) we have found that the detailed shape of this wing is not accurately captured by our broken power-law BLR model. 
This biases our estimate of the \sii\ fluxes, for which we need a different approach.
We therefore make a new, ad-hoc model for a small spectral region surrounding the \sii\ doublet ($\lala6680-6780$ rest-frame), consisting of a 5th-order polynomial for the BLR \ha\ wing, and two Gaussian components for each of the two \sii\ lines, representing the low and high-velocity parts of the doublet.
Ideally, we should apply this model to the spectrum integrated within an aperture which fully encloses the outflow material.
This can be done in PG\,1126-041, where we extract and model a series of spectra using various apertures within all the possible $R_{\rm out}$ (see Section \ref{ssec:energetics}), finding in all cases values for $n_{\rm e}$ within $\sim1000-2000\cmmc$.
MR\,2251-178, instead, presents a strong velocity gradient in the region of its outflow (see moment-1 maps in Fig.\,\ref{fig:kinematics}), which would artificially broaden the \sii\ lines when integrated over a large area.
Thus, in this galaxy we use a much smaller aperture ($0.05\arcsec$), which we centre at different locations within the outflow region, finding values for $n_{\rm e}$ within $\sim500-1000\cmmc$.
The intervals of $n_{\rm e}$ derived here are used in the Section below to compute the masses of the outflowing ionised gas.

To illustrate the described approach, in Fig.\,\ref{fig:SII} we show representative fits to the \sii\ doublets.
Clearly, both the \ha\ wings and the \sii\ lines are very well reproduced and the LV and HV-components are sufficiently separated.
There are cases, however, where the component separation is more ambiguous, but we stress that our results would not vary significantly if we used the entire lines, rather than the HV components alone, to derive $n_{\rm e}$.
We also note that the relation between \sii\ ratio and $n_{\rm e}$ of \citet{OF06}, which we adopt in our calculations, is perfectly compatible with the recent re-calibration by \citet{Kewley+19} for the range of line ratios that we find.

Given the emission lines available in our data, here we
have estimated the electron density using the `classical' method based on \sii\ ratios.
We note that the use of different density tracers \citep[such as the auroral or transauroral lines, e.g.][]{Holt+11} or of methods based on the ionisation parameter \citep[e.g.][]{BaronNetzer19} tend to output densities higher by about one order of magnitude \citep{Davies+20}, and consequently lead to lower ionised gas masses \citep[but see also][whose results suggests that these higher densities might be related to specific brighter and denser clumps that are not dominating the bulk of the outflow mass]{Mingozzi+19}.

\subsection{Outflow energetics}\label{ssec:energetics}
We now determine the physical properties of the kpc-scale outflow in MR\,2251-178 and PG\,1126-041.
In what follows, we assume that the outflow material in the QSOs is traced by the HV components discussed in Section \ref{ssec:kinematics}, and that it can be described as a collection of ionised gas clouds all having the same electron density $n_{\rm e}$\footnote{A variable cloud density implies a non-unitary multiplicative factor $\mathcal{C}\equiv\avg{n_{\rm e}}^2/\avg{n_{\rm e}^2}$ in eq.\,(\ref{eq:mout_ha}) and (\ref{eq:mout_oiii}). A constant $n_{\rm e}$ yields $\mathcal{C}\!=\!1$ \citep{CanoDiaz+12}}.

As before, we focus on the two brightest lines (\oiii\ and \ha) for which the HV components are more clearly visible, and use them as independent estimators for the outflow properties.
Following \citet{Cresci+17}, the mass of the outflowing ionised gas can be computed from the luminosity of the HV-\ha\ line $L_{\rm HV}^{{\rm H}\alpha}$ as
\begin{equation}\label{eq:mout_ha}
M_{\rm out}^{{\rm H}\alpha} = 3.2\times10^5 \left(\frac{L_{\rm HV}^{{\rm H}\alpha}}{10^{40}\ergs} \right) \left(\frac{100\cmmc}{n_{\rm e}}\right)\msun\,.
\end{equation}
The same quantity can be derived from the HV-\oiii\ line, following \citet[][see their Appendix B]{CanoDiaz+12}:
\begin{equation}\label{eq:mout_oiii}
M_{\rm out}^{\rm [OIII]} = 5.33\times10^4 \left(\frac{L_{\rm HV}^{\rm [OIII]}}{10^{40}\ergs} \right) \left(\frac{100\cmmc}{n_{\rm e}}\right) \frac{1}{10^{\rm[O/H]}}\msun
\end{equation}
where $10^{\rm [O/H]}$ is the oxygen abundance in Solar units.

The mass outflow rate, $\dot{M}_{\rm out}$, at a given radius $R_{\rm out}$ is derived using the simplified assumptions of spherical (or multi-conical) geometry and a constant outflow speed $v_{\rm out}$.
Following \citet{Lutz+20}, we have
\begin{equation}\label{eq:mout_dot}
\dot{M}_{\rm out} = \mathcal{H}\frac{M_{\rm out}v_{\rm out}}{R_{\rm out}} = 1.03\times10^{-9} \left(\frac{v_{\rm out}}{\kms}\right) \left(\frac{M_{\rm out}}{\msun}\right) \left(\frac{\kpc}{R_{\rm out}}\right)\mathcal{H}\msunyr
\end{equation}
where $\mathcal{H}$ is a multiplicative factor that depends on the adopted outflow history.
We assume a temporally constant $\dot{M}_{\rm out}$ during the flow time $R_{\rm out}/v_{\rm out}$, which leads to a radially decreasing density for the outflowing gas ($\rho\propto R^{-2}$, i.e., an `isothermal' case) and gives $\mathcal{H}\!=\!1$.
This choice agrees with the `time-averaged thin shell' approach \citep{Rupke+05b}, and has been extensively used in the literature to describe the energetics of the ionised and neutral phases of outflows \citep[e.g.][]{Arav+13,Heckman+15,Gonzalez-Alfonso+17,Veilleux+17}.
We note that some other works prefer to adopt $\mathcal{H}\!=\!3$ \citep[e.g.][]{Maiolino+12,Cicone+14,Feruglio+17}, corresponding to a constant volume density for the outflowing gas as resulting from a decaying outflow history where $\dot{M}_{\rm out}\!=\!0$ at the current time.
This is, however, at odds with the presence of UFOs in our sources.

The kinetic energy $E_{\rm kin}$ and luminosity $L_{\rm kin}$ (sometimes called `kinetic rate' and indicated as $\dot{E}_{\rm kin}$) are given by
\begin{equation}\label{eq:Ekin}
E_{\rm kin} = 9.94\times10^{42} \left(\frac{M_{\rm out}}{\msun}\right)\left(\frac{v_{\rm out}}{\kms}\right)^2 \erg\,
\end{equation}
\begin{equation}\label{eq:Lkin}
L_{\rm kin} = 3.16\times10^{35} \left(\frac{\dot{M}_{\rm out}}{\msunyr}\right)\left(\frac{v_{\rm out}}{\kms}\right)^2 \ergs\,
\end{equation}
and finally the momentum rate $\dot{p}_{\rm out}$ is computed as
\begin{equation}\label{eq:pdot}
\dot{p}_{\rm out} = 6.32\times10^{30} \left(\frac{\dot{M}_{\rm out}}{\msunyr}\right)\left(\frac{v_{\rm out}}{\kms}\right)\,{\rm dyn}\,.
\end{equation}

Eq.\,(\ref{eq:mout_ha})-(\ref{eq:pdot}) make use of different ingredients ($v_{\rm out}$, $R_{\rm out}$, $n_{\rm e}$, luminosity and metallicity), most of which can be determined using our model for the HV components.
In particular, we make use of `outflow velocity maps', derived as follows for the two lines in exam.
For a given Voronoi bin $k$, we define the outflow velocity $v_{\rm out}(k)$ as
\begin{equation}\label{eq:outflow_vel}
v_{\rm out}(k) = {\rm max}\,(|v_{5}(k)-v_{\rm sys}|\,,\,|v_{95}(k)-v_{\rm sys}|)
\end{equation}    
where $v_{5}$ and $v_{95}$ are, respectively, the 5-th and 95-th velocity percentiles of the modelled HV profile in that bin and $v_{\rm sys}$ is the bulk (or systemic) velocity of the galaxy, given by its newly determined redshift.
This definition is based on two assumptions.
The first reflects our ignorance on the true morphology of the outflow and on its orientation with respect to the observer: as projection effects are unknown, we make the ansatz that what best represents the intrinsic outflow speed is the velocity \emph{tail} of the profile, i.e., the $v_{5}$ or $v_{95}$ in eq.\,(\ref{eq:outflow_vel}).
These values are certainly better suited to represent $v_{\rm out}$ than the mean (or median) velocity of the line, which is strongly affected by projection effects and by dust absorption \citep[e.g.][]{Cresci+15}.
The second assumption is that the outflow material comes from the nuclear region, thus speeds must be computed with respect to the systemic velocity of the galaxy (the $v_{\rm sys}$ in eq.\,(\ref{eq:outflow_vel})) rather than to any other (local) reference frame.
This is justified by the fact that the HV component appears indeed to be confined within the central region, as shown in Fig.\,\ref{fig:kinematics}.

\begin{figure}
\begin{center}
\includegraphics[width=0.5\textwidth]{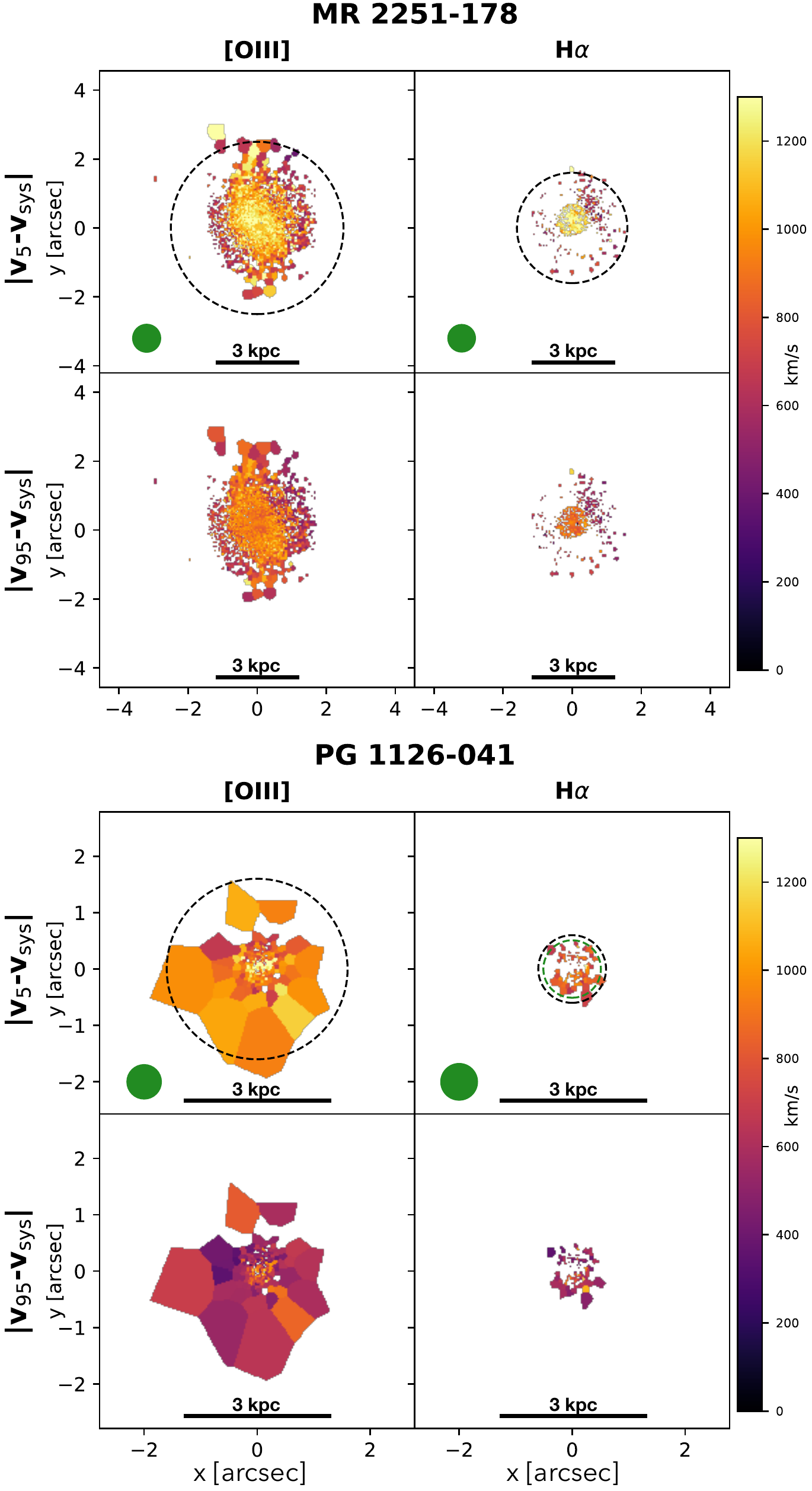}
\caption{\emph{Top panels:} \oiii\ and \ha\ velocity maps showing the 5th (first row) and 95th (second row) velocity percentiles of the HV components in MR\,2251-178. The dashed circles show the maximum $R_{\rm out}$ adopted for the calculation of the momentum boost. The green filled circles indicate the size of the PSF halo, which is markedly smaller than the spatial extent of the data. \emph{Bottom panels:} same for PG\,1126-041. The green dashed circle in the \ha\ panel shows the outflow radius corrected for the effect of the PSF.}
\label{fig:outflow_vel}
\end{center}
\end{figure}

Fig.\,\ref{fig:outflow_vel} shows maps for the two velocity terms that appear in eq.\,(\ref{eq:outflow_vel}) for the \oiii\ and \ha\ lines in MR\,2251-178 and PG\,1126-041.
As before, only HV components with peak S/N above $3$ are considered.
As expected for an outflow traced by optical emission lines, $|v_5-v_{\rm sys}|\!>\!|v_{95}-v_{\rm sys}|$ in virtually all spaxels, thus $v_{\rm out}$ is in practice given by the first term alone. 
Both systems show maximum $v_{\rm out}$ slightly above $1000\kms$.
Interestingly, in MR\,2251-178 $v_{\rm out}$ is larger at the centre and decreases at larger radii, which can be interpreted as due to projection effects in a spherically expanding wind.
In PG\,1126-041, instead, the kinematic pattern is less clear.
In general, the outflows are not perfectly circular and it is not straightforward to identify a characteristic outflow radius.
However, under the assumption of constant $\dot{M}_{\rm out}$ and $v_{\rm out}$, eq.\,(\ref{eq:mout_dot}) is expected to hold at any $R_{\rm out}$. 

The empty circles drawn in Fig.\,\ref{fig:outflow_vel} indicate the maximum possible outflow radius, $R_{\rm max}$, which we take equal to the maximum spatial extent of the HV component in the lines analysed.
Outflows in our QSOs are much more extended than the seeing-limited halos of the MUSE PSFs, which are shown as filled green circles with radius of $2R_{\rm halo}$ (see Appendix \ref{app:MUSE_PSF} for the core/halo decomposition of the PSFs) in Fig.\,\ref{fig:outflow_vel}. This indicates that the outflows are spatially resolved.
A correction such as $R'_{\rm max}=\sqrt{R^2_{\rm max}-R^2_{\rm halo}}$ yields $R'_{\rm max}\simeq R_{\rm max}$ in all cases, with the largest difference occurring for the \ha\ outflow of PG\,1126-041 (black vs green dashed circles in Fig.\,\ref{fig:outflow_vel}).

To calculate the quantities in eq.\,(\ref{eq:mout_ha})-(\ref{eq:pdot}), along with their associated uncertainty, we adopt a Monte-Carlo approach where we randomly extract $N\!=\!10^4$ possible $R_{\rm out}$ between $0.5\times{\rm PSF}_{\rm FWHM}$, a proxy for the minimum resolved spatial scale (taken from Table \ref{tab:resolution}), and $R_{\rm max}$.
The luminosity associated to a particular extraction is computed from the flux of the HV component within that aperture.
Similarly, for $v_{\rm out}$ we take a random velocity extracted within $R_{\rm out}$ from the maps shown in Fig.\,\ref{fig:kinematics}.
For $n_{\rm e}$ we assume a uniform distribution within the ranges determined in Section \ref{ssec:ne}, $500-1000\cmmc$ for MR\,2251-178 and $1000-2000\cmmq$ for PG\,1126-041.
For simplicity, we assume the metallicity to be about Solar, and we randomly sample it in the range $-0.3\!<\!{\rm[O/H]}\!<\!0.3$.
The \ha/\hb\ flux ratios that we derive in integrated spectra of the HV components are consistent with negligible dust extinction, thus no corrections for the estimated luminosities are required.

\begin{table*}
 \centering
% \begin{minipage}{140mm}
   \caption{Physical properties of the ionised outflow in MR\,2251-178 and PG\,1126-041 as derived from our analysis of the \ha\ and \oiii\ lines.}
   \label{tab:outflow}
   \setlength\tabcolsep{3.0pt}   
   \def\arraystretch{1.4}
   \begin{tabular}{l|ccccc|ccccc}
   \hline
   \hline
     line & \multicolumn{5}{c|}{\ha} & \multicolumn{5}{c}{\oiii} \\
     \hline
     parameter& $M_{\rm out}$ & $\dot{M}_{\rm out}$ & $E_{\rm kin}$ & $L_{\rm kin}/L_{\rm BOL}$ & $\dot{p}_{\rm out}$ & $M_{\rm out}$ & $\dot{M}_{\rm out}$ & $E_{\rm kin}$ & $L_{\rm kin}/L_{\rm BOL}$ & $\dot{p}_{\rm out}$\\  
     
     units & $10^6\msun$ & $\msunyr$ & $10^{56}\erg$ & $10^{-4}$ &  $(L_{\rm BOL}/c)$ & $10^6\msun$ & $\msunyr$ & $10^{56}\erg$ & $10^{-4}$ &  $(L_{\rm BOL}/c)$ \\  
     
     \hline
     MR\,2251-178 & $3.67^{+1.21}_{-0.85}$ & $4.29^{+5.36}_{-2.08}$ & $0.47^{+0.29}_{-0.21}$ & $3.69^{+7.03}_{-2.63}$ & $0.18^{+0.29}_{-0.11}$ & $3.41^{+2.49}_{-1.46}$ & $2.91^{+3.17}_{-1.54}$ & $0.39^{+0.38}_{-0.20}$ & $2.23^{+3.65}_{-1.59}$ & $0.11^{+0.16}_{-0.07}$\\
     PG\,1126-041 & $0.37^{+0.14}_{-0.12}$ & $0.97^{+1.06}_{-0.39}$& $0.04^{+0.06}_{-0.02}$ & $1.91^{+8.76}_{-1.15}$ & $0.12^{+0.27}_{-0.06}$ & $0.52^{+0.19}_{-0.17}$ & $0.60^{+0.47}_{-0.22}$ & $0.05^{+0.03}_{-0.02}$ & $1.10^{+1.77}_{-0.57}$ & $0.07^{+0.08}_{-0.03}$\\
     \hline
   \end{tabular}
% \end{minipage}
\end{table*}

With all the ingredients at our disposal, we then compute eq.\,(\ref{eq:mout_ha})-(\ref{eq:pdot}) $N$ times, take the medians of the resulting distributions as representative values for our measurements, and use the 84th and 16th percentiles for their upper and lower uncertainties.
Table \ref{tab:outflow} summarises the outflow properties in MR\,2251-178 and PG\,1126-041, including our estimates for the kinetic efficiency $L_{\rm kin}/L_{\rm BOL}$, and for the so-called `momentum boost', that is the ratio between the momentum flux of the outflowing gas $\dot{p}_{\rm out}$ and the momentum flux initially provided by radiation pressure from the AGN ($L_{\rm BOL}/c$, where $L_{\rm BOL}$ is taken from Table \ref{tab:properties}).
Despite the limitations associated to these measurements, both the \ha\ and the \oiii\ lines give compatible values for the properties of the outflow in MR\,2251-178 and PG\,1126-041.
In all cases, mass outflow rates are modest ($0.6-4.3\msunyr$) and the kinetic efficiency is small ($1-4\times10^{-4}$).
These values are in good agreement with the $\dot{M}_{\rm out}-L_{\rm BOL}$ and $L_{\rm kin}-L_{\rm BOL}$ scaling relations reported by \citet{Fiore+17}, which corroborates the validity of our method and indicates that MR\,2251-178 and PG\,1126-041 are similar to the bulk of the galaxy population featuring AGN-driven ionised winds.

We discuss further the implication of these findings on the physical mechanisms associated to the AGN wind in Section \ref{ssec:physics}.
We stress that our analysis accounts only for the ionised component of the outflow, and it is well possible that the presence of a neutral (atomic and/or molecular) component may significantly increase the overall outflow mass and energy budget \citep[e.g.][]{Cicone+14,Fluetsch+19}.

\section{Discussion}\label{sec:discussion}
\subsection{On the origin of the nebula around MR\,2251-178} \label{ssec:nebula}
As known from several previous studies \citep[e.g.][]{Bergeron+83,Macchetto+90,Shopbell+99}, MR\,2251-178 is surrounded by a a huge emission-line nebula, with a total size of $\sim200\kpc$ and ionised gas mass of $\sim6\times10^{10}\msun$.
Early kinematic estimates based on the \ha\ and \oiii\ lines indicate that the extended regions of the nebula counter rotate with respect to the regions immediately surrounding the QSO \citep{Norgaard-Nielsen+86,Shopbell+99}, suggesting the presence of (at least) two distinct origins for the ionised gas around MR\,2251-178.
Different origins have been proposed for the extended nebula, including tidal debris from an interaction with a companion, debris from a major merging event, outflow from the QSO, cooling flow, or photo-ionisation of a large \hi\ envelope.
The latter option seems to be in better agreement with the regular large-scale kinematics suggested by the data.
Unfortunately, interferometric \hi\ data are not available for this source, and only a tentative \nai\ absorption feature is visible in our MUSE data, too faint to be modelled.

\begin{figure}
\begin{center}
\includegraphics[width=0.5\textwidth]{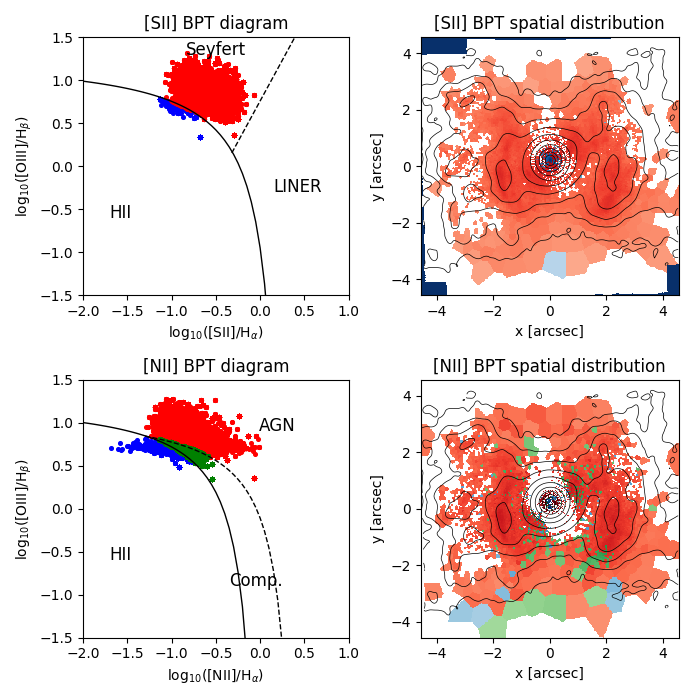}
\caption{\emph{Left panels}: spatially resolved \sii- and \nii- BPT diagrams for MR\,2251-178. Each dot corresponds to a single Voronoi bin. Red (blue) dots represent regions whose excitation is dominated by star formation (AGN), green dots are used for composite (in the \nii-BPT) or LI(N)ER (in the \sii-BPT) excitation regimes. Solid and dashed black lines show the separation between the different regimes, from \citet{Kewley+01,Kewley+06} and \citet{Kauffmann+03}.
Only lines with S/N$>2$ are used.
\emph{Right panels}: \sii- and \nii- BPT maps, color coded according to the dominant excitation regime. Darker colors indicate higher \sii\ (top panel) or \nii\ (bottom panel) intensities.
Iso-intensity contours show the integrated \oiii\ line, smoothed to a FHWM resolution of $0.25\arcsec$.
The ionisation of the nebula is vastly dominated by the AGN.
}
\label{fig:bpt_mr2251}
\end{center}
\end{figure}

As our MUSE data cover only the innermost $\sim12\times12\kpc^2$ region, they cannot provide information on how the central regions are related to the outer parts.
However, it is possible to confirm the counter-rotating nature of the inner nebula: we find red-shifted (blue-shifted) motion towards the east (west), in contrast with the large-scale kinematics found by \citet{Shopbell+99}.
Both the \sii\ and the \nii\ BPT diagrams, shown in Fig.\,\ref{fig:bpt_mr2251}, indicate that the AGN is by far the primary excitation driver for the nebula, although star formation may be relevant in the very centre and a few composite regions are present.

As discussed in Section \ref{ssec:kinematics}, in spite of the fact that the ionised gas is structured in a series of shells, the low velocity dispersion and the orientation of the velocity gradient in the LV component (see Fig.\,\ref{fig:kinematics}) do not favour an origin as gas outflowing from a past nuclear activity.
Also, while the velocity gradient may recall that of a regularly rotating disc, the fact that velocities in the red-shifted region drop down to zero beyond a few kpc from the center indicates that the inner nebula is not made by cold gas in circular orbits within the gravitational potential of the host. 
Instead, the complex morphology and kinematics of this nebula closely resemble those of the extended emission line regions (EELRs) around fading AGNs studied by \citet{Keel+15}, who concluded that photo-ionisation of tidal debris from minor mergers or flybys are the most likely origins for these structures.

Clearly, a detailed study of the metallicity of the ionised nebula is key to understand its origin.
We attempted a first-order metallicity calculation using the procedure of \citet{Storchi-Bergmann+98}, which makes use of the \nii/\ha\ and the \oiii/\hb\ ratios alone and is specifically designed for the diffuse gas in active galaxies. 
Unfortunately, this resulted in a positive [O/H] gradient ranging from Solar-like values at the center of the nebula to $\sim10$ times Solar at $R\simeq5\kpc$, which we find unrealistic.
We believe it is more likely that the gradients in line ratios are due to radial variations in the ionization field rather than in the metal content.
Interestingly, though, some numerical models of major mergers predict positive metallicity gradients for the merger remnant, as high-$Z$ gas is moved inside-out by the enhanced AGN feedback activity \citep[e.g.][]{Cox+06}.
As MUSE wide field mode observations of this system are currently ongoing, we leave a more detailed study of the small and large-scale metallicity structure of the nebula - along with its extended kinematics - to a future work.

\subsection{Physical mechanisms for the wind propagation}\label{ssec:physics}
Relating the momentum boost of these kpc-scale outflows to that derived for the sub-pc scale high-speed ($\sim0.1c$) wind traced by X-ray offers fundamental clues on the physical mechanism associated to the wind propagation.

Following the theoretical model of \citet{King10}, the X-ray wind is pushed by radiation pressure from the AGN and shock-heats the surrounding gas, inflating a hot bubble.
Inverse-Compton is the primary cooling process at work, and the bubble evolution depends on the cooling timescale.
For short cooling timescales, most of the bubble energy is radiated away and the expanding (cold) shell communicates with the ambient medium only via its ram pressure, transferring its momentum flux to it (`momentum-driven' regime).
Long timescales imply that the bubble keeps expanding while conserving its energy (`energy-driven' regime), efficiently sweeping up a considerable portion of the ISM and driving outflows of several thousands of $\msunyr$.
This simple model may not capture additional effects due to the non-spherical geometry \citep{ZubovasNayakshin14} or to the multi-phase nature of the wind \citep{FCQ12}, but has the considerable advantage of predicting a $M_{\rm BH}-\sigma$ relation in broad agreement with the observations.
%Also, winds may evolve with time and transit from one regime to the other.} 

We already determined in Section \ref{ssec:energetics} the momentum boosts for the galaxy-scale ionised outflows in MR\,2251-178 and PG\,1126-041.
We now use X-ray measurements from the literature to determine the sub-pc scale energetics in a similar manner.
Following \citet{NardiniZubovas18} we assume that the UFO is launched from the escape radius $r_{\rm esc}\equiv2GM_{\rm BH}/v^2_{\rm UFO}$, i.e., the radius at which the observed outflow speed v$_{\rm UFO}$ corresponds to the escape velocity from the black hole, and write the mass outflow rate as
\begin{equation}\label{eq:mout_ufo}
    \dot{M}_{\rm UFO} \simeq 0.3\left(\frac{\Omega}{4\pi}\right)\left(\frac{N_{\rm H,UFO}}{10^{24}\cmmq}\right)\left(\frac{M_{\rm BH}}{10^8\msun}\right)\left(\frac{v_{\rm UFO}}{c}\right)^{-1}\msunyr
\end{equation}
where $\Omega$ is the solid angle subtended by the wind, $N_{\rm H,UFO}$ is the gas column density and $v_{\rm UFO}$ is the outflow speed.
We take $M_{\rm BH}$, $v_{\rm UFO}$ and $N_{\rm H,UFO}$ from Table \ref{tab:properties}, and assume fiducial values of $\Omega/4\pi$ between $0.1$ and $0.2$, roughly corresponding to the detection rate of UFOs per source per observation in local AGN samples \citep[e.g.][]{Tombesi+10}.
UFO momentum rates $\dot{p}_{\rm UFO}$ are determined using a formula analogous to eq.\,(\ref{eq:pdot}): we find $0.66^{+1.24}_{-0.43}/(L_{\rm BOL}/c)$ for PG\,1126-041 and $\gtrsim0.19/(L_{\rm BOL}/c)$ (lower limit) for MR\,2251-178.

\begin{figure}
\begin{center}
\includegraphics[width=0.48\textwidth]{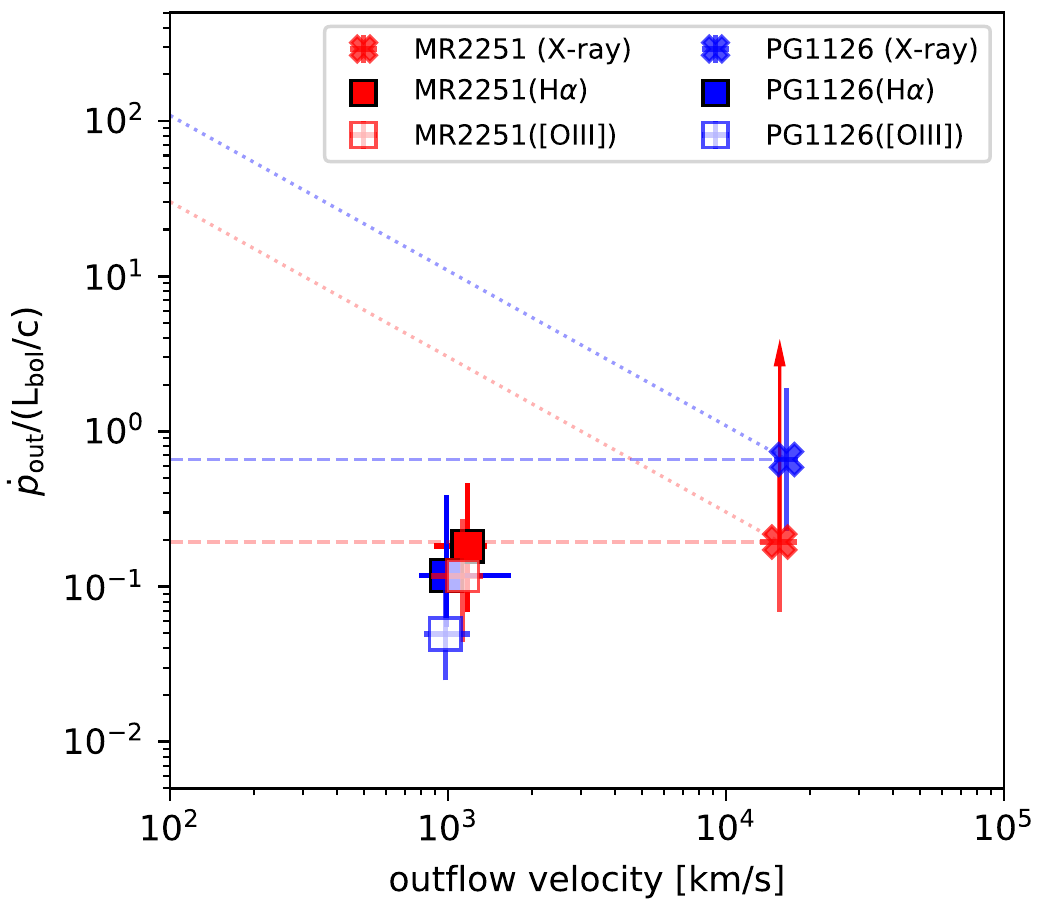}
\caption{Momentum boost for MR\,2251-178 (red markers) and PG\,1126-041 (blue markers) as a function of the outflow velocity. Crosses show the high-speed, sub-pc scale wind observed in X-ray by \citet{Reeves+13} and \citet{Giustini+11} (see text). Filled and empty squares are used for the kpc-scale ionised outflow from the \ha\ and \oiii\ lines respectively, as derived in this work. Dashed horizontal (dotted diagonal) lines show predictions for a momentum-driven (energy-driven) wind.}
\label{fig:momentum_boost}
\end{center}
\end{figure}

Fig.\,\ref{fig:momentum_boost} shows the momentum-boost as a function of the outflow speed for MR\,2251-178 and PG\,1126-041.
Prediction for momentum conserving (energy conserving) winds are shown with dashed horizontal (dotted diagonal) lines originating from the X-ray data points.
Error-bars on the UFO momentum rates are large, especially in the case of MR\,2251 where only a rough lower limit can be established.
While these uncertainties are difficult to overcome, our measurements indicate that the momentum rate of the kpc-scale outflow is roughly compatible with that of the sub-pc scale wind, favouring a `simple' momentum-driven scenario for these two galaxies.
We note that, given the measured outflow velocities, an energy-driven scenario would be favoured only if the momentum fluxes were underestimated by about two orders of magnitude.
Unless the AGN-driven outflow in MR\,2251-178 and PG\,1126-041 is completely dominated by a massive component of molecular gas, we find it unlikely that a perfectly energy-driven mechanism is at work here.

However, the hypothesis of massive molecular outflows associated to these systems is not unrealistic.
Various studies \citep[e.g.][]{Carniani+15,Fiore+17,Bischetti+19,Fluetsch+19} have compared molecular and ionised outflow rates in different AGN-hosting galaxies as a function of their $L_{\rm BOL}$. 
Given the typical $L_{\rm BOL}$ of our galaxies (some $10^{45}\ergs$), outflow rates for an hypothetical molecular component may indeed exceed those measured for the ionised gas by a factor of $\sim100$. 
Unfortunately the correlation between the different outflow phases is very uncertain, and only with dedicated molecular gas observations (e.g. with ALMA) a more complete picture of the outflow properties in these systems can be achieved.

\subsection{Comparison with other QSOs}\label{ssec:other_sources}
While the galaxy-scale (ionised) outflows in MR\,2251-178 and PG\,1126-041 appear to be propelled by momentum-driven winds, one may wonder whether the same occurs in other QSOs.
In order to answer this question, we gathered from the literature high-quality measurements for other $8$ well-studied QSOs, including a lensed system at $z\simeq3.9$ (APM\,08279+5255), hosting both UFOs at sub-pc scales and galaxy-scale molecular/atomic outflows.
For each of these sources we have re-computed the UFO mass outflow rate (and the wind energetics, subsequently) using eq.\,(\ref{eq:mout_ufo}), starting from known estimates for $M_{\rm BH}$, $N_{\rm H,UFO}$ and $v_{\rm UFO}$.
All estimates for molecular outflows have been re-scaled to the same luminosity-to-mass conversion factor $\alpha_{\rm CO}=0.8\,({\rm K}\kms\pc^2)^{-1}\msun$, typical of QSOs, starburst and submillimeter galaxies \citep{DownesSolomon98,CarilliWalter13,Bolatto+13}.
This allowed us to build a small, yet reliable high-quality sample for which the outflow properties have been determined in an homogeneous way. 
References for each system are given in Appendix \ref{app:other_sources} and the detailed properties of pc- and galaxy-scale winds resulting from this analysis are listed Table \ref{tab:other_sources}.
Below, instead, we present our main results.

\begin{figure*}
\begin{center}
\includegraphics[width=1.0\textwidth]{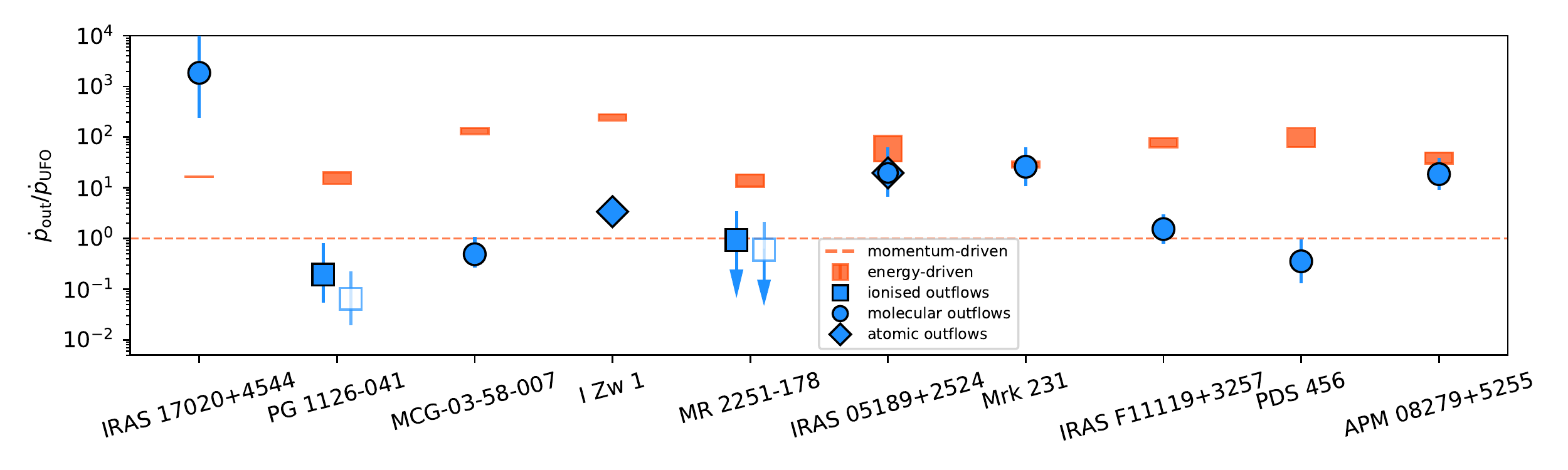}
\caption{Ratio between the galaxy-scale and sub-pc scale outflow momentum rates for different QSOs hosting UFOs. Measurements for individual galaxies are shown as squares (ionised outflows), circles (molecular outflows) and diamonds (atomic outflows) with error-bars. Systems are ordered by increasing $L_{\rm BOL}$. The horizontal dashed line shows the prediction for momentum-driven winds ($\dot{p}_{\rm out}/\dot{p}_{\rm UFO}\!=\!1$), individual predictions for energy-driven winds are shown as orange rectangles. Our \ha\ and \oiii\ measurements for MR\,2251-178 and PG\,1126-041 are shown separately using filled and empty symbols, respectively.}
\label{fig:momentum_boost_ratio}
\end{center}
\end{figure*}

Fig.\,\ref{fig:momentum_boost_ratio} shows the ratio between the momentum rate of the galaxy-scale outflow ($\dot{p}_{\rm out}$) and that of the pc-scale wind ($\dot{p}_{\rm UFO}$) for all galaxies in our sample, and compares it to predictions for momentum-driven or energy-driven regimes (horizontal dashed line and orange rectangles, respectively).
We reiterate that the $\dot{p}_{\rm out}$ values used for MR\,2251-178 and PG\,1126-041 refer to the ionised gas, and not to molecular/neutral outflow as in the other sources considered.
Remarkably, with the exception of IRAS\,17020+4544\footnote{for which $\dot{p}_{\rm out}$ is very uncertain, see discussion in \citet[][Section 3.1]{Longinotti+18}}, all measurements for $\dot{p}_{\rm out}/\dot{p}_{\rm UFO}$ seems to be compatible with one regime or the other, with no system being markedly located at an intermediate position.
This demonstrates that the two theoretical regimes proposed, in spite of their simplicity, capture very well the physics of the AGN wind propagation.

However, the interpretation of Fig.\,\ref{fig:momentum_boost_ratio} is not straightforward and two possible scenarios can be discussed.
A possibility is that, at any given time, galaxy-scale winds are either in a momentum-driven or in an energy-driven regime, transiting from one regime to the other when certain physical conditions are met.
Within the framework described by \citet{KingPounds15}, this transition may occur when central black holes have grown sufficiently enough to comply with the $M_{\rm BH}-\sigma$ relation. 
When this happens, feedback transits from a poorly efficient (momentum-driven) phase, where the black hole still accretes material at a fast pace, to a highly efficient (energy-driven) regime, which causes the ejection of most of the ISM out of the galaxy, eventually quenching the black hole growth and star formation processes altogether.
In this framework, the lack of intermediate regimes shown by Fig.\,\ref{fig:momentum_boost_ratio} would indicate that the transition occurs on short timescales.
Clearly, independent measurements for $M_{\rm BH}$ and $\sigma$ would be needed to test this hypothesis. 

The scenario proposed, however, is not supported by most recent theoretical models of AGN feedback \citep[e.g.][]{Richings+18,Costa+20}, which predict energy-driven galaxy-scale outflows in virtually all cases.
In this context, observed sub energy-driven regimes would be attributed to an inefficient coupling between the UFO and the ISM, causing only part of the kinetic energy of the former to be transferred to the latter.
As a possible counter-argument to this interpretation, we note the bimodality shown by the QSOs in Fig.\,\ref{fig:momentum_boost_ratio} implies a coupling `efficiency factor' that is either $\sim1$ or very close to whatever value is required by the momentum-driven regime, which is somewhat difficult to justify.
However, we stress that the observed bimodality does not arise in similar studies on different QSO samples \citep[e.g.][]{Smith+19}.
A homogeneous, multi-phase study of a larger sample of QSOs  hosting both sub-pc and galaxy-scale winds and featuring a wide range of $L_{\rm BOL}$ is required to confirm the trend shown by Fig.\,\ref{fig:momentum_boost_ratio}.

\subsection{Timescales}\label{ssec:timescales}
An important caveat associated to momentum boost arguments is that they implicitly assume that the QSO radiation force ($L_{\rm BOL}/c$) powering the pc-scale UFO is the same as that powering the galaxy-scale wind.
However, flow timescales $\tau_{\rm flow}\!\equiv\!R_{\rm out}/v_{\rm out}$ for galaxy-scale outflows are of the order of the $\Myr$, while $L_{\rm BOL}$ can substantially vary on much sorter times. 
Therefore, comparing the energetics of the two components is done under the assumption that the $L_{\rm BOL}$ measured at the current epoch, which powers the UFO, does not differ dramatically from the $L_{\rm BOL}$ averaged over the entire duration of the flow.
In this context, the exceptionally high $\dot{p}_{\rm out}$ of IRAS\,17020+4544 can be interpreted as due to a much higher AGN luminosity at previous epochs, when the molecular outflow was launched.
In fact, this is the only QSO in our sample for which $\tau_{\rm flow}$ ($\sim2\Myr$) is significantly smaller than the time that the UFO would take to inject a total amount of kinetic energy equivalent to that of the large-scale wind ($\tau_{\rm kin}\equiv E_{\rm kin}/L_{\rm kin,UFO}\sim60\Myr$), which implies that the UFO and the molecular outflow that we observe in this system are not causally connected.
Yet, the fact that all the other sources in Fig.\,\ref{fig:momentum_boost_ratio} are compatible with the predicted regimes suggests that the underlying assumption of approximately constant $L_{\rm BOL}$ over the flow time is physically motivated, at least for the targets presented here.
Analogous considerations apply to the case where the UFO duty cycle is much shorter than the flow time.

\section{Conclusions}\label{sec:conclusions}
Galaxy wide winds produced by AGN are routinely observed around galaxies at all redshifts at which AGN are observed.
They are thought to have a fundamental impact on galaxy evolution, in particular by quenching star formation in high-mass systems.
Despite substantial observational efforts, the physical mechanism by which the wind propagates from the (sub-pc scale) nuclear region to the (kpc scale) interstellar medium, eventually leading to galaxy scale outflows, is still elusive.
Progress in this direction can only be achieved by studying the properties of the wind on different physical scales \citep[e.g.][]{Tombesi+15,Feruglio+15,Veilleux+17,Tombesi+17,Smith+19,Bischetti+19,Mizumoto+19,Gaspari+20}.

To pursue this methodology, in this work we have used MUSE NF-mode data to study the properties of the ionised gas around two systems, MR\,2251-178 ($z\!=\!0.064$) and PG\,1126-041 ($z\!=\!0.060$), which host massive, sub-pc scale, ultra-fast ($\sim0.1c$) outflows (UFOs) visible via blue-shifted Fe K absorption line in their nuclear X-ray spectrum \citep{Gofford+11,Giustini+11}.
Our MUSE data cover a region of about $12\times12\kpc^2$ around the nucleus, and have sensitivity, spatial and spectral resolutions that are optimal to study the kinematics of the ionised gas from the \ha\ and \oiii\ emission lines.
We employed a method based on a multi-component analysis of the line profiles, which allowed us to disentangle the large-scale emission of the diffuse gas from the nuclear (unresolved) emission of the BLR, as the latter contaminates the whole field of view due to the extended wings of the MUSE AO-assisted PSF.
Emission from diffuse gas is further decomposed into a high-velocity (HV) component, which traces the large-scale, more turbulent outflowing material, and a low-velocity (LV) component, which exhibits a more regular and quiet kinematics.

Our results can be summarised as follows.
\begin{itemize}
    \item In both systems the HV component is spatially concentrated within a few kpc from the centre and features a high velocity dispersion (up to $\sim800\kms$) and a mean blue-shifted speed of a several tens $\kms$ in MR\,2251-178, or a few hundreds $\kms$ in PG\,1126-041.
    Undoubtedly, this component traces a nuclear outflow driven by the AGN.
    \item The LV component in MR\,2251-178 is structured in a series of concentric shells distributed in the east-west direction, tracing the inner regions of the giant ($200\kpc$-wide) ionised nebula that surrounds this galaxy. Its kinematics show a clean velocity gradient, misaligned with respect to the orientation of the shells, and a low ($\sim80\kms$) velocity dispersion. We speculate that tidal debris from recent minor mergers or flybys can explain the observed properties of this structure.
    \item In LV component in PG\,1126-041 is likely associated to the regularly rotating disc of the galaxy, which does not seem to be perturbed by the nuclear activity.
    \item Outflows are characterised by small mass rates ($0.6-4.3\msunyr$) and poor kinetic efficiencies ($1-4\times10^{-4}$).
    \item The momentum rates associated to the large-scale \ha\ and \oiii\ outflows are compatible with those inferred from the sub-pc scale measurements in the X-rays, in agreement with a momentum-driven wind propagation. Pure energy-driven winds are excluded unless our outflow rates are underestimated by about two orders of magnitude, which may occur in the presence of a massive molecular wind.
    \item We have compared the outflow energetics of MR\,2251-178 and PG\,1126-041 with those of a small sample of well-studied QSOs from the literature, each hosting both sub-pc scale UFOs and molecular and/or atomic winds on galaxy scales. We have found that, in all systems but one, winds are either in a momentum-driven or in an energy-driven regime. Uncertainties associated to the variability of the UFOs and to the UFO-to-ISM coupling efficiency prevent us from establishing whether or not the observed bimodality is intrinsic to the physics of AGN-driven winds.
\end{itemize}

A caveat associated to this study is the lack of information on atomic or molecular gas, which prevents us from determining in full the outflow energetics in the two QSOs in exam.
Follow-up observations (e.g., with ALMA) are essential to obtain a more complete picture of the outflow properties in these systems and to provide useful constraints to AGN-driven wind theories.

\begin{acknowledgements}
AM and GC acknowledge the support by INAF/Frontiera through the "Progetti Premiali" funding scheme of the Italian Ministry of Education, University, and Research. EN acknowledges financial support from the agreement ASI-INAF n.2017-14-H.0. EP acknwoledges financial support from ASI/INAF contract 2017-14-H.0 and from PRIN MIUR project “Black Hole winds and the Baryon Life Cycle of Galaxies: the stone-guest at the galaxy evolution supper”.

\end{acknowledgements}

%-------------------------------------------------------------------

\bibliographystyle{aa} % style aa.bst 
\bibliography{ufo} % your references aa.bib

%-------------------------------------------------------------------
%-------------------------------------------------------------------
\appendix
\section{The AO-assisted PSF of MUSE NFM}\label{app:MUSE_PSF}
\begin{figure*}
\begin{center}
\includegraphics[width=0.9\textwidth]{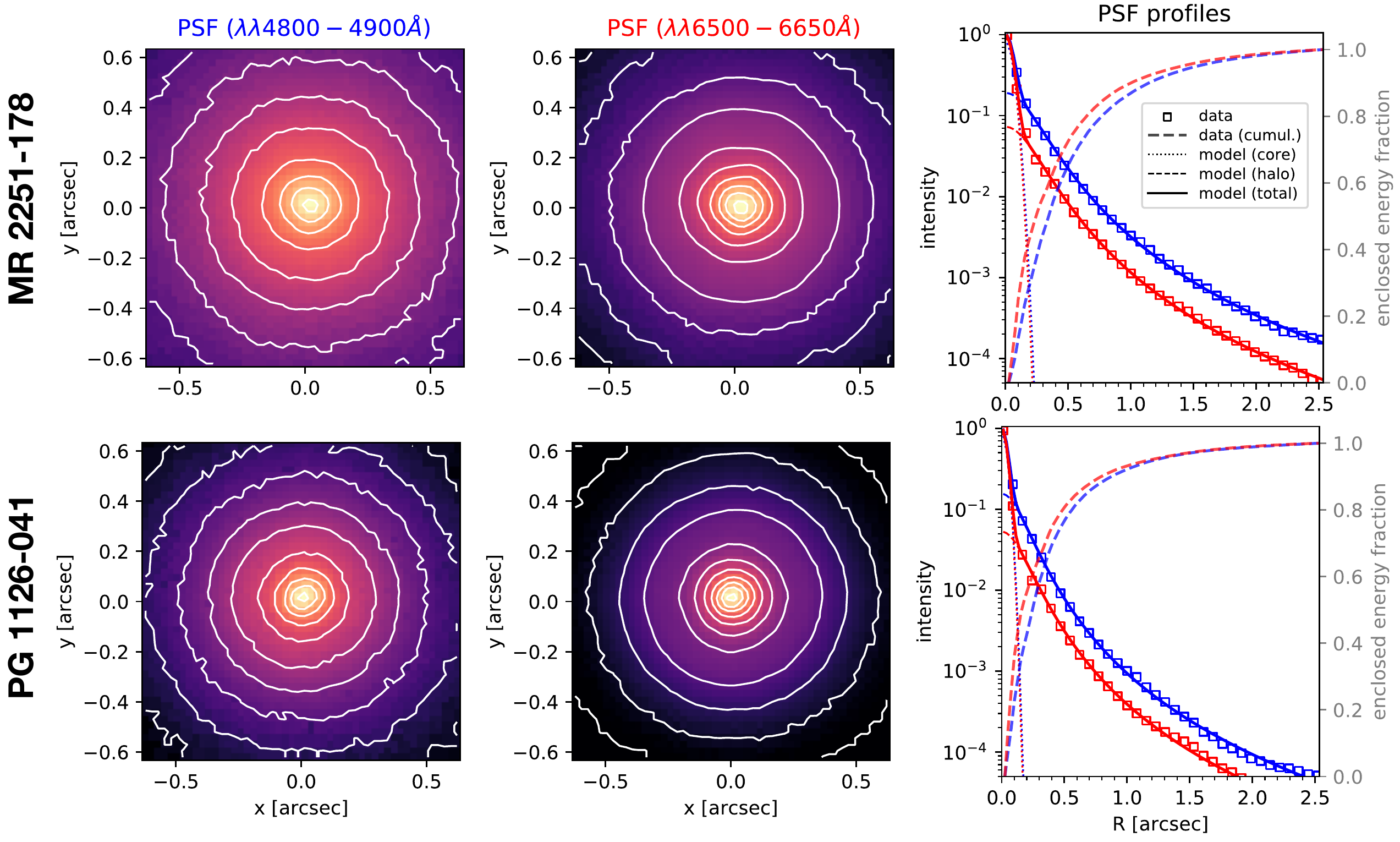}
\caption{MUSE AO-assisted PSF for MR\,2251-178 (\emph{top panels}) and PG\,1126-041 (\emph{bottom panels}) as traced by the intensity of the BLR emission. The leftmost and central panels show separate PSF maps for the blue regime ($\lala4800-4900\AA$) and for the red regime ($\lala6500-6650\AA$). Each map is normalised to its peak value. Iso-contours are the same for all maps and spaced by a factor of $2$. The rightmost panels show the azimuthally averaged PSF profiles (empty squares, one data-point every three is plotted for a better clarity) and the enclosed energy fraction (thick dashed lines). The solid lines show our best-fit models, given by the sum of a Gaussian core (dotted lines) and a Moffat halo (thin dashed lines).}
\label{fig:PSF_MUSE}
\end{center}
\end{figure*}

In this Appendix we describe the MUSE NFM PSFs derived from our modelling of the BLRs in two different spectral regimes, as discussed in Section \ref{ssec:blr}.
The PSF maps, obtained by integrating the BLR models over the rest-frame range $\lala4800-4900\AA$ (the `blue' regime) and $\lala6500-6650\AA$ (the `red' regime), are shown in Figure \ref{fig:PSF_MUSE}, together with their azimuthally-averaged profiles.
The full width at half maximum (FWHM) of the PSF profile in MR\,2251-178 (PG\,1126-041) is $76$ ($46$) mas for the red regime and $106$ ($68$) mas for the blue regime.
However, most of the light falls far beyond such distances: in MR\,2251-178 (PG\,1126-041) half of the total energy is enclosed within a diameter of $440$ ($270$) mas in the red and $640$ ($440$) mas in the blue.
These estimates are reported in Table \ref{tab:resolution} (along with the MUSE velocity resolution in the red and blue regimes) and clearly indicate that, as expected, the AO works better at longer wavelengths, leading to a more centrally concentrated PSF (i.e., a higher Strehl ratio). 
Also, the MUSE data for PG\,1126-041 have better angular resolution that those of MR\,2251-178, in agreement with the superior seeing conditions during the observations ($\sim0.48\arcsec$ vs $\sim0.84\arcsec$).
Overall, the fact that our BLR-traced PSFs agree with expectations from AO-assisted observational techniques is an excellent validation of our BLR modelling scheme and, in general, a promising starting point for our analysis.

We further decompose the PSF profiles into the sum of a core and a halo component, which we model with a Gaussian and a Moffat profile, respectively.
This model has four free parameters (one for the core, two for the halo, plus the normalisation) which we fit to the data to determine their optimal values.
In all cases we find a unique and well defined minimum $\chi^2$ in the parameter space\footnote{We have also attempted a fit using two Moffat functions, but found that such model is highly degenerate.}.
Our best-fit model gives a very good description of the PSF profiles, as shown in the rightmost panels of Figure \ref{fig:PSF_MUSE}.

We define core and halo `radii', $R_{\rm core}$ and $R_{\rm halo}$, as the standard deviations of the Gaussian and Moffat components.
These values are used to compare the PSFs with the spatial extent of the outflows in Section \ref{ssec:energetics} and are reported at the bottom of Table \ref{tab:resolution}, along with the fraction of the total light enclosed within the core, $f_{\rm core}$.

\begin{table}
 \centering
% \begin{minipage}{140mm}
   \caption{Main PSF properties and velocity resolution of our MUSE data for two different spectral ranges, $\lala4800-4900\AA$ (`blue') and $\lala6500-6650\AA$ (`red').}
   \label{tab:resolution}
   \begin{tabular}{l|cc|cc|l}
   \hline\hline
   \multicolumn{6}{c}{MUSE-NFM spatial and spectral resolution}\\
   \hline
                        & \multicolumn{2}{c|}{MR\,2251-178} & \multicolumn{2}{c|}{PG\,1126-041} &\\
    \hline
                        & blue & red & blue & red  &\\
    \hline
    (1) PSF$_{\rm FWHM}$ & $106$ & $76$  & $68$    & $46$&mas  \\
    (2) PSF$_{\rm D50}$ & $640$ & $440$ & $440$   & $270$ &mas\\
    (3) $\sigma_{\rm MUSE}$ & $65$  & $50$  & $65$    & $50$&$\kms$ \\ 
    \hline
    \multicolumn{6}{c}{core/halo PSF decomposition}\\
    \hline
    (4) $R_{\rm core}$ & $52$ & $48$ & $39$ & $37$ & mas\\
    (5) $R_{\rm halo}$ & $393$ & $387$ & $297$ & $317$ & mas\\
    (6) $f_{\rm core}$ & $0.14$ & $0.31$  &  $0.18$  &  $0.38$  &\\ 
    \hline
    \noalign{\vspace{5pt}}
    \multicolumn{6}{p{0.42\textwidth}}{\textbf{Row description.} (1) FWHM of the MUSE PSF; (2) Diameter enclosing half of the total PSF energy; (3) minimum resolved velocity dispersion; (4)-(5) core and halo radii, defined as the standard deviations of the best-fit Gaussian and Moffat components; (6) fraction of energy enclosed within the core.}\\
   \end{tabular}
% \end{minipage}
\end{table}

\section{Supplementary material}\label{app:other_sources}
\begin{table*}
 \centering
   \caption{Main physical properties for the QSOs studied in Section \ref{ssec:other_sources}, and for their pc-scale and galaxy-scale winds.}
   \label{tab:other_sources}
   \setlength\tabcolsep{3.0pt}   
   \def\arraystretch{1.3}
   \begin{tabular}{l|cc|ccc|ccc}
   \hline
    \hline
            & \multicolumn{2}{c|}{AGN properties} & \multicolumn{3}{c|}{X-ray wind} & \multicolumn{3}{c}{Galaxy-scale wind} \\
    \hline
    parameter & $\log(L_{\rm BOL}/\lsun)$ & $\log(M_{\rm BH}/\msun)$ & $v_{\rm UFO}$ & $N_{\rm H,UFO}$ & $\dot{p}_{\rm UFO}$ & $E_{\rm kin}$ & $\dot{p}_{\rm out}$ & tracer \\
    units &   &   & $c$  & $10^{24}\cmmq$ & $(L_{\rm BOL}/c)$ & $10^{56}\erg$ & $(L_{\rm BOL}/c)$ & \\
     \hline
IRAS\,05189+2524  & $12.22$ & $8.50^{+0.37}_{-0.46}$ & $0.11\pm0.01$ & $0.27^{+0.23}_{-0.12}$ & $0.32^{+0.60}_{-0.22}$ & $37.3^{+5.7}_{-4.3}$ & $6.35\pm1.03$ & OH+Na \\
I\,Zw\,1  & $12.01$ & $6.97^{+0.06}_{-0.07}$  & $0.26\pm0.01$ & $0.60\pm0.15$ & $0.12^{+0.04}_{-0.03}$ & $6.2^{+1.8}_{-1.4}$ & $0.40^{+0.14}_{-0.10}$ & Na \\
PDS\,456  & $13.61\pm0.10$ & $9.24\pm0.17$ & $0.25\pm0.01$ & $0.69\pm0.09$ & $0.64^{+0.38}_{-0.24}$ & $6.4^{+9.6}_{-3.8}$ & $0.23^{+0.35}_{-0.14}$ & CO \\
Mrk\,231  & $12.60\pm0.10$ & $8.38\pm0.09$ & $0.067\pm0.008$ & $0.27^{+0.36}_{-0.15}$ & $0.11^{+0.16}_{-0.06}$ & $12.8$ & $2.77^{+0.71}_{-0.56}$ & CO,OH \\
IRAS\,F11119+3257  & $12.67\pm0.10$ & $8.29\pm0.15$ & $0.25\pm0.01$ & $6.4^{+0.8}_{-1.3}$ & $1.72^{+1.08}_{-0.67}$ & $522^{+311}_{-214}$ & $2.66^{+0.96}_{-0.72}$ & CO,OH \\
IRAS\,17020+4544  & $[11.13,11.79]$ & $7.59\pm0.08$ & $0.09$ & $^a0.05^{+0.36}_{-0.05}$ & $0.05^{+0.32}_{-0.04}$ & $51.0$ & $89.2^{+60.5}_{-35.9}$ & CO \\
MCG-03-58-007  & $[11.62,11.95]$ & $8.34\pm0.05$ & $0.075\pm0.01$ & $0.78\pm0.32$ & $1.70^{+1.06}_{-0.77}$ & $0.57\pm0.03$ & $0.81^{+0.60}_{-0.28}$ & CO \\
$^b$APM\,08279+5255  & $14.24$ & $10.1\pm0.4$ & $0.18\pm0.04$ & $0.35^{+0.35}_{-0.17}$ & $0.16^{+0.36}_{-0.11}$ & $179$ & $2.82$ & CO \\
\hline
\noalign{\vspace{5pt}}
    \multicolumn{9}{p{1.0\textwidth}}{\textbf{Notes.} UFO mass rates are computed using eq.\,(\ref{eq:mout_ufo}) with $(\Omega/4\pi)$ in the range $[0.1,0.2]$, with the exceptions of I\,Zw\,1 and PDS\,456 for which covering fractions of $~0.5$ are directly measured. 
    $^a$ The original entry from \citet{Longinotti+15} is $\log(N_{\rm H,UFO}/\cmmq)=23.99^{+0}_{-1.86}$, which we interpret as a truncated Gaussian distribution. $^b$ A magnification of $\mu=4$ is assumed.}
   \end{tabular}
\end{table*}

Table \ref{tab:other_sources} lists the main parameters used in the calculation of the UFO and the galaxy-scale outflow energetics for the QSOs studied in Section \ref{ssec:other_sources}.
We furnish the appropriate references for each system below.

\emph{IRAS\,05189+2524}. 
We estimate the black hole mass from the $M_{\rm BH}-\sigma$ relation \citep{KH13}, using as a range the two values of $\sigma$ determined by \citet[][$137\kms$]{Dasyra+06} and \citet[][$265\kms$]{Rothberg+13}. 
$L_{\rm BOL}$ is taken from \citet{Veilleux+13}, while the UFO velocity and column density from \citet{Smith+19}.
The galaxy-scale outflow is seen via different tracers: CO \citep{Lutz+20}, OH \citep{Gonzalez-Alfonso+17} and Na \citep{Rupke+17}. 
The OH and Na phases dominate the energetics, and we sum them for the purpose of determining the global outflow properties.

\emph{I\,Zw\,1}.
In this system $M_{\rm BH}$ has been accurately measured by \citet{Huang+19} via a dedicated reverberation mapping campaign.
We take $L_{\rm BOL}$ from \citet{Veilleux+13} and the properties of the X-ray wind from \citet{ReevesBraito19}, who also determined a large value for the wind covering factor (here we assume $\Omega/4\pi=0.5$).
A CO wind is undetected in this QSO. 
Instead, there is evidence for a neutral outflow \citep{Rupke+17}, which we use in our calculations.

\emph{PDS\,456}.
We use the $M_{\rm BH}$ determined by \citet{Nardini+15} using the $\lambda5100\AA$ luminosity and the \hb\ FWHM.
We take the averaged $L_{\rm BOL}$ from two different estimates, the first from the $\lambda5100\AA$ luminosity with bolometric correction \citep{Richards+06}, the second using the $\lambda1350\AA$ luminosity from \citet{Hamann+18}.
The properties of the X-ray UFO come from \citet{Nardini+15}, where they specifically determined a wind opening angle $\Omega/4\pi$ of $0.5$.
A galaxy-scale outflow is detected in CO by \citet{Bischetti+19} and consists of an extended component, in turn partitioned into four different blobs, and a compact component, which largely dominates the energetics of the system. 
For the purpose of measuring the outflow energetics with consider both components together.

\emph{Mrk\,231}.
As for PDS\,456, $M_{\rm BH}$ is derived using the $\lambda5100\AA$ luminosity and the \hb\ FWHM \citep[from][]{NardiniZubovas18}.
We take $L_{\rm BOL}$ from \citet{Veilleux+13}, and the X-ray wind velocity and $N_{\rm H,UFO}$ from \citet{Feruglio+15}.
The molecular (CO) outflow energetics come from \citet{Lutz+20}, who used data originally from \citet{Cicone+12}.
A large-scale outflow is also observed in OH \citet{Gonzalez-Alfonso+17}, but with properties that are compatible with those of the wind detected in CO. 
Following \citet{NardiniZubovas18}, we assume the CO-based estimates as representative for the large-scale outflow.

\emph{IRAS\,F11119+3257}.
Also for this system $M_{\rm BH}$ comes from the $\lambda5100\AA$ luminosity and the \hb\ FWHM \citep[from][]{NardiniZubovas18}, and $L_{\rm BOL}$ from \citet{Veilleux+13}.
X-ray wind properties are from \citet{Tombesi+15}.
Molecular outflows are observed in both OH \citep{Tombesi+15} and CO \citep{Veilleux+17} but on different physical scales, with the former being confined within $1\kpc$ from the center while the latter being visible up to larger ($\sim15\kpc$) distances.
It is therefore likely that the two molecules trace distinct outflow episodes.
For simplicity we show here only the CO energetics, but that determined for the OH is compatible within the quoted uncertainties.

\emph{IRAS\,17020+4544}.
We infer the black hole mass from the $M_{\rm BH}-\sigma$ relation of \citet{KH13} using the \oiii\ velocity dispersion \citep[$125\kms$][]{WangLu01} as a proxy for the stellar $\sigma$ .
We bracket the value of $L_{\rm BOL}$ using two measurements, the first from the $\lambda5100\AA$ \citep{Hao+15} and the second (more conservative) from that assumed in \citet{Longinotti+15}, from which we take also the X-ray wind properties.
The properties of the CO molecular outflow are from \citet{Lutz+20}.
In \citet{Longinotti+18}, the outflow properties of this QSO were found to be compatible with an energy-driven regime (although with large uncertainties).
The differences here are due to a factor of $\sim5$ lower covering factor assumed for the UFO and to a factor $\sim20$ lower gas column density, which arises as we interpret the entry $\log(N_{\rm H,UFO}/\cmmq)=23.99^{+0}_{-1.86}$ in Table 2 of \citet{Longinotti+15} as a truncated Gaussian distribution, leading to a median $\log(N_{\rm H,UFO}/\cmmq)\simeq22.73$. 

\emph{MCG-03-58-007}.
As before, $M_{\rm BH}$ is inferred from the $M_{\rm BH}-\sigma$ using the \oiii\ velocity dispersion \citep[$185\kms$,][]{Braito+18} as a proxy for $\sigma$.
The $L_{\rm BOL}$ is bracketed by the two values quoted in \citet{Braito+18}, determined using either \oiii\ or IR luminosities.
From the same work we take the properties of the X-ray wind, for which we consider only the slowest of the two components observed, which is the only one persisting in different observations.
The properties of the CO outflows are taken from \citet{Sirressi+19}.

\emph{APM\,08279+5255}.
Unlike the other systems considered, this is a lensed QSO at $z\simeq3.9$. 
While lens models can not constrain very accurately the value of the magnification $\mu$, on which most quantities relevant to this analysis depend, indirect methods indicate $\mu\!=\!4$ as the best possible choice \citep{Saturni+18}. We consider this value for the magnification and take $M_{\rm BH}$ and $L_{\rm BOL}$ from \citet{Saturni+18}.
The X-ray wind properties come from \citet{Chartas+09}, using only the slowest component (see their Table 3, model $8$, absorber $1$) which was also identified separately by \citet{Hagino+17}.
The properties of the CO molecular outflow are taken from \citet{Feruglio+17}, using their model with $\mu=4$.

\end{document}